\documentclass[a4paper,fleqn,usenatbib]{mnras}

\usepackage[T1]{fontenc}
\usepackage{ae,aecompl}

\usepackage{graphicx}	
\usepackage{amsmath}	
\usepackage{amssymb}	

\usepackage{makecell}
\usepackage{lscape}
\usepackage{color}
\usepackage[figuresright]{rotating}

\def\mean#1{\left< #1 \right>}
\newcommand*{\p}{\turnbox{12}{$\,'$}\;}

\definecolor{green2}{rgb}{0.13, 0.55, 0.13}
\definecolor{gray2}{rgb}{0.6, 0.6, 0.6}
\definecolor{orange}{rgb}{1.0, 0.49, 0}

\title[Central galaxy alignments]{Intrinsic alignments in redMaPPer clusters -- I. Central galaxy alignments and angular segregation of satellites}

\author[Huang et al.]{
Hung-Jin Huang$^{1}$\thanks{E-mail: hungjinh@andrew.cmu.edu},
Rachel Mandelbaum$^{1}$\thanks{E-mail: rmandelb@andrew.cmu.edu},
Peter E. Freeman$^{1, 2}$, 
Yen-Chi Chen$^{1, 2, 3}$, \newauthor
\ Eduardo Rozo$^{4},$ 
Eli Rykoff$^{5}$ \&
Eric J. Baxter$^{6}$
\vspace{0.1in}\\
$^{1}$McWilliams Center for Cosmology, Department of Physics, Carnegie Mellon University, Pittsburgh, PA 15213, USA\\
$^{2}$Department of Statistics, Carnegie Mellon University, Pittsburgh, PA 15213, USA\\
$^{3}$Department of Statistics, University of Washington, Seattle, WA 98195, USA\\
$^{4}$Department of Physics, University of Arizona, 1118 E. Fourth St., Tucson, AZ 85721, USA\\
$^{5}$SLAC National Accelerator Laboratory, Menlo Park, CA 94025, USA \\
$^{6}$Center for Particle Cosmology, Department of Physics, University of Pennsylvania, Philadelphia, PA 19104
}

\date{Accepted XXX. Received YYY; in original form ZZZ}

\pubyear{2016}

\begin{document}
\label{firstpage}
\pagerange{\pageref{firstpage}--\pageref{lastpage}}
\maketitle

\begin{abstract}

The shapes of cluster central galaxies are not randomly oriented, but rather exhibit coherent alignments with the shapes of their parent clusters as well as with the surrounding large-scale structures.
In this work, we aim to identify the galaxy and cluster quantities that most strongly predict the central galaxy alignment phenomenon among a large parameter space with a sample of 8237 clusters and 94817 members within $0.1<z<0.35$, based on the redMaPPer cluster catalog constructed from the Sloan Digital Sky Survey.
We first quantify the alignment between the projected central galaxy shapes and the distribution of member satellites, to understand what central galaxy and cluster properties most strongly correlate with these alignments. Next, we investigate the angular segregation of satellites with respect to their central galaxy major axis directions, to identify the satellite properties that most strongly predict their angular segregation.
We find that central galaxies are more aligned with their member galaxy distributions in clusters that are more elongated and have higher richness, and for central galaxies with larger physical size, higher luminosity and centering probability, and redder color. Satellites with redder color, higher luminosity, located closer to the central galaxy, and with smaller ellipticity show a stronger angular segregation toward their central galaxy major axes. 
Finally, we provide physical explanations for some of the identified correlations, and discuss the connection to theories of central galaxy alignments, the impact of primordial alignments with tidal fields, and the importance of anisotropic accretion.

\end{abstract}





\section{Introduction}
\label{sec:intro}

In the framework of the standard cold dark matter (CDM)-dominated Universe, cosmic structures grow
hierarchically. Small galaxies form first, then merge and group together through channels of the
filamentary network to form clusters of galaxies \citep{White78,Blumenthal84}.  During the process
of structure formation, the distribution and orientation of galaxies may be set by the surrounding
gravitational tidal fields, or be disturbed by activities such as mergers or feedback processes due
to supernova or active galactic nuclei. In this work, we refer to any net preferred orientation toward some
reference direction or any existing galaxy shape correlations caused by these physically-induced
events as {\em intrinsic} alignments (in contrast with the coherent alignments induced by
gravitational lensing). For recent
reviews, see \citet{Joachimi15}, \citet{Kiessling15} and \citet{Kirk15}.

Intrinsic alignments occur on a variety of
scales.  
On large scales, several Mpc and above, galaxies show a net tendency to align radially towards
overdensities
\citep[e.g.,][]{Mandelbaum06,Hirata07,Okumura09,Joachimi11},
and more detailed analysis of the cosmic web indicates coherent alignments  along the stretching direction
of filaments \citep{Tempel15,Chen15b,Rong16}.  One of the leading theoretical models for
intrinsic alignments at large scales ($\gtrsim$ 6 Mpc) is the linear alignment model, which relates
the alignment strength linearly to the smoothed tidal field at the time of
galaxy formation \citep{Catelan01,Hirata04a}, or variations of that model that include nonlinear
evolution of the density field \citep{Bridle07,Blazek15}.  Based on a sample of luminous red
galaxies (LRGs), \citet{Singh15} adopted the above alignment models to quantify the large-scale
alignment amplitude as a function of several LRG properties. They found that the alignment amplitude
becomes stronger toward more luminous LRGs residing in higher mass halos \citep[see also][]{Hirata07,Joachimi11}.

On small scales, within galaxy clusters, there are two types of alignments. The first type is the
alignment of satellite major axes toward the center of their host dark matter (DM) halo, for which
the observational proxy usually is the brightest cluster galaxy (BCG). 
This is often called {\em satellite} (or {\em radial}) alignment. 
Satellite alignment is believed to originate from the tidal
torque induced primarily from the gravitational field of the DM halo
\citep{Ciotti94,Kuhlen07,Pereira08,Faltenbacher08,Tenneti15}. Observationally, the existence of satellite
alignment is still controversial, with \citet{Pereira05,Agustsson06,Faltenbacher07,Singh15} reporting
detections of the signal, while \citet{Hung12,Schneider13,Chisari14,Sifon15} found no significant
detection.  Some of this tension may arise from selection effects, as discussed by \cite{Singh15}.  
In addition, \citet{Hao11} cautioned about the possibility of spurious 
satellite alignment signals due to systematic errors (the contamination from the
diffuse light from BCGs).  We will report our measurement of satellite alignment in red-sequence
Matched-filter Probabilistic Percolation (redMaPPer)
clusters and present detailed systemic analysis in the upcoming Paper II. 
In the current paper, we focus on the second type of alignment, called {\em central galaxy} alignment.

Central galaxy alignment refers to the tendency of the major axis of the central galaxy to align with that of its host DM
halo, for which the observational signature is that satellites (which we use as a tracer of the DM
halo shape) preferentially reside along the central's major axis direction.  
This type of alignment is also termed ``{\em BCG} alignment'' in the literature, as it is
  often assumed that the brightest galaxy within each cluster is the central galaxy (the central
  galaxy paradigm, see \citealt{vdBosch05}). However, \citet{Skibba11} showed that 40\%, and
  \citet{Hoshino15} that 20-30\%, of BCGs are not the galaxies that are located closest to the
  center of the cluster potential well. The fact that the redMaPPer algorithm  identifies centrals not
  only based on their luminosity but also on their color and local galaxy density enables us to
  select a more robust set of central galaxies for our intrinsic alignment study. Therefore,
  through out this work, we will use the term ``central galaxy alignment'' for our result, and keep
  the term ``BCG alignment'' when referring to previous works that utilize the BCG as a proxy for
  the central galaxy.

Unlike satellite alignment, the observational evidence
for central galaxy alignment is strong and uncontroversial \citep[e.g.,][]{Sastry68, Binggei82, Niederste-Ostholt10},
and it can be explained by two possible
physical mechanisms. The first is the filamentary nature of matter
accretion \citep{Dubinski98}, and the second is primordial alignment with the tidal field set by
both the host dark matter halo and large-scale structure \citep{Faltenbacher08}.
Since central galaxy alignment is robustly detected with existing large datasets, many
studies have investigated its dependence on physical predictors such as
central galaxy luminosity, color, host halo mass, redshift, and so on, in order to better understand the
physical origin of the effect 
\citep{Brainerd05,Yang06,Azzaro07,Faltenbacher07,Wang08,Siverd09,Agustsson10,Niederste-Ostholt10,Hao11}.
There is general agreement that the central galaxy alignment signal is stronger for red and luminous centrals,
and shows higher significance when using red satellites as tracers.  However, some controversies
still remain about the importance of other predictors besides luminosity or color. Furthermore, some
of the 
previous studies started with the assumption that only a few predictors could be important in
determining 
the central galaxy alignments, and therefore performed an analysis based only on those predictors without considering others,
ignoring potential degeneracies among predictors when splitting and comparing subsamples.

In Paper I, our goal is to present a comprehensive analysis of the predictors of central galaxy alignments. We
include as many physical properties as possible, and properly account for potential correlations
among them with the help of a linear regression analysis.  We also discuss potential systematic effects based on
signals obtained from various shape measurement methods. 
The two main questions we aim to address are 1) what central and cluster properties are the strongest
predictors of the strength of central galaxy alignments? 2) What kinds of satellites are more likely to lie
along the major axis direction of their host centrals?  
We build corresponding linear regression models, 
use variable selection techniques to select important predictors, and
further quantify their significance. 
Finally, we discuss possible physical origins for these
selected predictors and compare our result with the literature. 

The paper is organized as follows. In Sec.~\ref{sec:data}, we describe our data and definitions
of the physical quantities used in the linear regression analysis. Details of the linear regression
process are described in Sec.~\ref{sec:LR}. Sec.~\ref{sec:result}
presents our measurement of central galaxy alignment and results of the variable selection process. Sec.~\ref{sec:shape-align} discusses the detected central galaxy alignment signal
for three different shape measurement methods, and the interpretation of those findings. The physical origins of our identified featured predictors for central galaxy alignments with the cluster shape and angular segregation of satellites with
respect to the central galaxy major axis are discussed in detail in Secs.~\ref{sec: BCG origin}
and~\ref{sec: satellite origin}, respectively.
We conclude and summarize our key findings in Sec.~\ref{sec:summary}.

Throughout this paper, we adopt the standard flat $\Lambda$CDM cosmology with $\Omega_m=0.3$ and
$\Omega_{\Lambda}=0.7$. All the length and magnitude units are presented as if the Hubble constant
were 100~km~s$^{-1}$~Mpc$^{-1}$.  
In addition, we use $\log$ as shorthand for the 10-based logarithm,
and $\ln$ for the natural logarithm.



\section{Data and measurements}
\label{sec:data}

In this section, we introduce the data that we analyze in this work, including the definitions of the galaxy
cluster and galaxy properties that we use.  All data used in this paper came 
from the Sloan Digital Sky Survey (SDSS) I/II surveys   The SDSS I
\citep{York2000} and II surveys imaged roughly $\pi$ steradians
of the sky, and followed up approximately one million of the detected
objects spectroscopically \citep{Eisenstein2001,Richards2002,Strauss2002}. The imaging was carried
out by drift-scanning the sky in photometric conditions
\citep{Hogg2001,Ivezic2004}, in five bands
($ugriz$) \citep{Fukugita1996,Smith2002} using a
specially-designed wide-field camera
\citep{Gunn1998}. These imaging data were used to create
the catalogs that we use in this paper.  All of
the data were processed by completely automated pipelines that detect
and measure photometric properties of objects, and astrometrically
calibrate the data \citep{Lupton2001,Pier2003,Tucker2006}. The SDSS-I/II imaging
surveys were completed with a seventh data release
\citep{Abazajian2009}, but we use the processed data from an 
improved data reduction pipeline that was part of the eighth data
release, from SDSS-III \citep{Aihara11}; and an improved
photometric calibration \citep[`ubercalibration',][]{Padmanabhan2008}.

\subsection{Galaxy cluster catalog}

We use member galaxies in the redMaPPer v5.10 cluster catalog\footnote{http://risa.stanford.edu/redmapper/} to study galaxy alignments in 
galaxy clusters. The redMaPPer cluster catalog is constructed based on photometric galaxy
samples with a magnitude cut $m_i < 21.0$ from the SDSS data release
eight (DR8; \citealt{Aihara11}) over a total area of $\sim$10,000 deg$^{2}$.  Details of the
redMaPPer cluster finding algorithm and
properties of the SDSS redMaPPer catalogs can be found in 
\citet{Rykoff14,Rozo14,Rozo15a,Rozo15b}.  Briefly, the redMaPPer algorithm has two
stages: the red-sequence calibration, and the cluster-finding stage. With a set of red spectroscopic
galaxies as training sample, redMaPPer first constructs a redshift-dependent evolutionary red-sequence model,
including zero-point, tilt, and scatter. The calibrated red-sequence
model is then used to group red galaxies at similar redshifts into clusters, assuming certain radial
and luminosity filters.

One of the features of the redMaPPer algorithm is that it is probabilistic, which enables users to
select suitable samples to do statistics.  For each cluster, it provides the central galaxy
probability, $P\rm_{cen}$, for the top five potential BCGs, and all potential member galaxies are
assigned with a membership probability, $p\rm_{mem}$, according to their color, magnitude, and
position information. The photometric redshift $z$ for each cluster is estimated from
high-probability members; and the cluster richness, $\lambda$, is defined by summing the
membership probabilities over all cluster members.

In this work, we restrict our analysis to clusters with richness $\lambda \ge 20$, corresponding
to a halo mass threshold of $M\rm_{200m} \gtrsim 10^{14}\ h^{-1} M_{\odot}$ \citep{Rykoff12}, and 
photometric redshift in the range $0.1 \leq z \leq 0.35$. The lower redshift
limit is selected so as to minimize edge effects from the training sample when doing calibration,
while the upper redshift cutoff is set such that the sample of clusters is volume-limited
\citep{Rykoff14}. All in all, there are 10702 clusters within this redshift and richness range.
To perform higher quality statistics, we only explore satellite galaxies with
membership probability $p\rm_{mem} \geq 0.8$ when doing linear regression analysis, and restrict to
satellites with $p\rm_{mem} \geq 0.2$ when defining cluster shape (while weighting those satellites
appropriately by their values of $p\rm_{mem}$).

\subsection{Definitions and measurements of physical parameters}
\label{subsection:definitions}

In this subsection, we describe many of the physical parameters that we
will use to study central galaxy alignments.

\subsubsection{Galaxy ellipticity}

The galaxy ellipticity used for the majority of this work is corrected for the effect of the PSF
using the re-Gaussianization shape measurement method
(see Sec.~\ref{sec:shape} for detail).  We use the components of the distortion $e_1$ and $e_2$ \citep{Bernstein02}
provided from the \citet{Reyes12} (or R12) and \citet{Mandelbaum05} (or M05) catalogs by fitting the 'atlas images'
\citep{Stoughton02} in both $r$ and $i$ bands.  The distortion can be related to the axis ratio $b/a$ as
\begin{equation}\label{eq:e1e2}
(e_1,\ e_2)=\frac{1-(b/a)^2}{1+(b/a)^2}(\rm{cos}\ 2\alpha,\ \rm{sin}\ 2\alpha),
\end{equation}
where $\alpha$ is the position angle of the major axis. The total galaxy distortion $e$ is calculated as
\begin{equation}\label{eq:e}
e=\sqrt{e_1^2+e_2^2}
\end{equation}

\subsubsection{Galaxy alignment angles}

Once the galaxy position angle is known, we can
assign each satellite its central galaxy alignment angle, $\theta\rm_{cen}$, and satellite alignment angle, $\phi\rm_{sat}$.
  
\begin{figure*}
\begin{center}
\includegraphics[trim=0.15cm 0cm 1.0cm 0cm,width=0.7\textwidth]{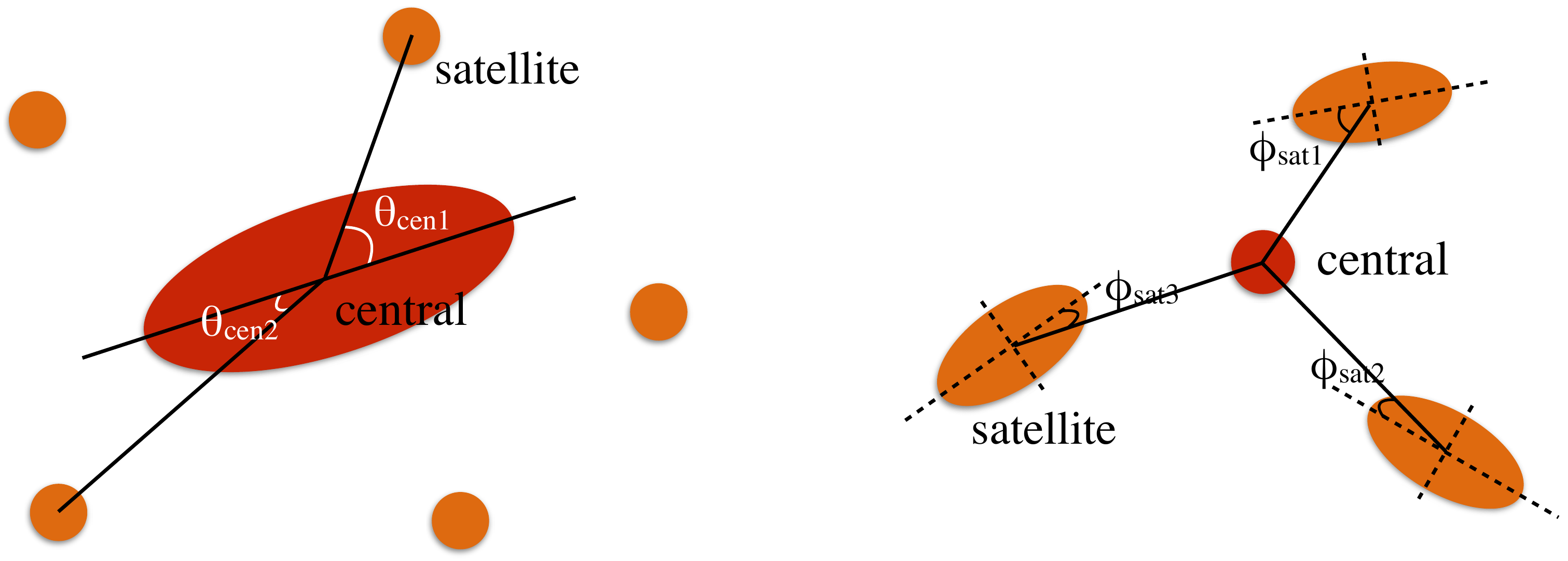}
\caption{Illustration of the galaxy alignment angles. The left panel shows the definition of central
  alignment angle $\theta\rm_{cen}$, while the right panel shows the definition of satellite
  alignment angle $\phi\rm_{sat}$.}
\label{fig:illustration}
\end{center}
\end{figure*}

The central galaxy alignment angle $\theta\rm_{cen}$ is defined as the angle between the major axis of the central galaxy and
the line connecting the central to the satellite galaxy, as illustrated in the left panel of
Fig.~\ref{fig:illustration}. Calculating $\theta\rm_{cen}$ requires a viable shape
measurement for the central galaxy (but not the satellites). Within the redshift range $0.1 \leq z
\leq 0.35$, there are 8237 centrals with shape measurements in the R12 catalog, resulting in 94817
central-satellite pairs with satellites that have $p\rm_{mem} \geq 0.8$.

The satellite alignment angle $\phi\rm_{sat}$ is defined as the angle between the major axis of the 
satellite galaxy and the line connecting its center to the central, as shown in the right panel of
Fig.~\ref{fig:illustration}. Calculating $\phi\rm_{sat}$ requires a shape measurement for
the satellite galaxy. In this paper, we only consider $\phi\rm_{sat}$ as a
potential predictor of central galaxy alignments; future work will include a detailed analysis of satellite
alignments.

We restrict both $\theta\rm_{cen}$ and $\phi\rm_{sat}$ to the range [0$^\circ$,
90$^\circ$] due to symmetry. By definition, $\theta\rm_{cen}=0^\circ/90^\circ$ indicates a satellite
located 
along the major/minor axis of the central. A satellite is radially/tangentially aligned with the central if
$\phi\rm_{sat}=0^\circ/90^\circ$.

\subsubsection{Cluster position angle and ellipticity}
\label{subsec:cluster e}

We follow the method used in \citet{Niederste-Ostholt10} to define the orientation and ellipticity
of the redMaPPer clusters from their satellite distributions. In order to have enough member
galaxies to trace the shape of each cluster, we use all member galaxies with membership probability
$p\rm_{mem}\ge 0.2$. We calculate the reduced second moments from the positions of member
galaxies, weighted by $p\rm_{mem}$:
\begin{equation} \label{eq:Mxx}
M_{\rm xx} \equiv \mean{\frac{x^2}{r^2}} = \frac{\sum\limits_{i} p_{{\rm mem},i} \frac{x_i^2}{r_i^2} }{ \sum\limits_{i} p_{{\rm mem},i}}
\end{equation}
and likewise for $M_{\rm yy}$ and $M_{\rm xy}$; by definition, $M_{\rm xx}+M_{\rm yy}=1$.  Here $x_i$ is the distance of member galaxy $i$ from the cluster
center. 
We can then define the cluster ellipticity as 
\begin{equation}\label{eq:QU}
(Q,\ U) = \frac{1-b/a}{1+b/a}(\rm{cos}\ 2\beta,\ \rm{sin}\ 2\beta) = (M_{\rm xx}-M_{\rm yy},\ 2M_{\rm xy}),
\end{equation}
where $b/a$ is the cluster projected minor-to-major axis ratio and $\beta$ is the cluster position angle (P.A.). The
cluster ellipticity can then be calculated via 
\begin{equation}
{\rm cluster\ } e=\sqrt{Q^2+U^2}.
\end{equation}

With the $1/r^2$ weighting (an explicitly spherically-symmetric weight function) in the reduced second moments, the derived cluster ellipticity tends
to be underestimated.  We show later that this does not change our conclusion regarding how cluster
ellipticity affects  the central galaxy alignment. 

\subsubsection{Central galaxy dominance}

The central galaxy dominance parameter is defined as the difference in the r-band absolute magnitude of the central galaxy
and the mean magnitude of the first and second brightest satellites:
\begin{equation}\label{eq:dom}
{\rm Central\ dominance}\  \equiv\ {\rm Central}\ ^{0.1}M_{\rm r} - \frac{^{0.1}M_{\rm r,1st}+^{0.1}M_{\rm r, 2nd}}{2}.
\end{equation}

We calculate the central galaxy dominance parameter using only $p\rm_{mem} \geq 0.8$ members. For the very few
clusters (134 out of 8237) that have only one member satisfying the $p_{\rm mem} \geq 0.8$
criterion, we simply use the difference between the absolute magnitudes of central and that 
member galaxy to define the central galaxy dominance.
Smaller central dominance values correspond to more dominant central galaxies.

\subsubsection{Central galaxy probability}
For each cluster, the redMaPPer catalog contains the five most likely central galaxy candidates, each
with  centering probability $P\rm_{cen}$. In this paper, we use the most probable
central as our central galaxy, and measure the central galaxy and satellite alignment angles of the associated central-satellite pairs. Over 80\% of our centrals have $P\rm_{cen} \geq 0.7$.

\subsubsection{Galaxy absolute magnitude}

We calculate the absolute magnitude for each galaxy using the luminous red galaxy (LRG) templates in
the {\tt kcorrect} package (v4.2) distributed by \citet{Blanton07}. The {\tt kcorrect} software
determines the best composite fit to the observed galaxy spectral energy distribution (SED) with the
\citet{Chabrier03} initial mass function (IMF) and a variety of \citet{Bruzual03} stellar population
synthesis models differing in star formation histories and metallicities.
We use extinction-corrected SDSS model magnitudes and the photometric redshift $z$ provided in redMaPPer as input, and k-correct the magnitudes of all galaxies in our sample to $z=0.1$. 

\subsubsection{Galaxy effective radius}

The effective radius we report in this paper is the circularly-averaged half-light radius, defined as
\begin{equation}\label{eq:Reff}
R_{\rm eff}\ \equiv\ \sqrt{\frac{b}{a}} \,R_{\rm deV},
\end{equation}
where b/a is the semi-minor to semi-major axis ratio taken from the SDSS parameter {\tt
  deVAB\_r}, and $R\rm_{deV}$ is the semi-major half-light radius, {\tt
  deVRad\_r}. Both parameters are estimated as part of the SDSS DR8 pipeline by  fitting de Vaucouleurs light profiles to
galaxy r-band images. Here we convert the value of
$R_{\rm deV}$ from the provided angular units to physical units ($h^{-1}$kpc), using the redshift
$z$ of the host cluster.

\subsubsection{Member distance from the cluster center}

For each satellite galaxy, we compute its projected distance, $r$, to the central galaxy, to check for radial dependence in the central galaxy alignment signal. 
To fairly compare among satellite galaxies in clusters with different halo masses, we further
normalize $r$ by the estimated halo radius, $R\rm_{200m}$, corresponding to the radius within which the average density of the enclosed mass is 200 times the mean density, $\overline{\rho}$. 
We first use the mass-richness relation provided in Eq.~B4 of \citet{Rykoff12},
\begin{equation}\label{eq:Mhalo}
\ln{\left(\frac{M_{\rm 200m}}{h^{-1}_{70}\ 10^{14}\ {\rm M_{\odot}}}\right)} = 1.72 + 1.08\ \ln{\frac{\lambda}{60}},
\end{equation}
to estimate $M\rm_{200m}$. Then we compute $R\rm_{200m}$ via the definition $M_{\rm 200m} =
(4\pi/3)\  200\overline{\rho}\ R_{\rm 200m}^{3} $.
Our conclusions would not change even if applying different mass-richness relations recently calibrated via weak lensing \citep{Simet16} or via clustering of clusters \citep{Baxter16}.

\subsubsection{Cluster member concentration $\Delta_{\rm R}$}

Recently, \citet{Miyatake16} found that the average projected distance of member galaxies from the cluster center, defined as 
\begin{equation}\label{eq:Rmem}
\overline{R}_{\rm mem} = \frac{\sum\limits_{i}p_{{\rm mem},i}R_i}{\sum\limits_{i}p_{{\rm mem},i}},
\end{equation} 
not only describes the concentration of the member galaxy distribution in the cluster, but also
plays a role in determining the large-scale clustering of redMaPPer clusters at fixed mass. Here
$p_{{\rm mem},i}$ is the membership probability of the $i$-th member galaxy, and $R_i$ is the
physical separation between that galaxy and its corresponding cluster central galaxy.

To properly model the richness and redshift dependence in $\overline{R}_{\rm mem}$, we use
another parameter, $\Delta_{\rm R}$, defined in Eq.~22 of \citet{Baxter16} as an indicator of
cluster member concentration at fixed $\lambda$ and $z$:
\begin{equation}\label{eq:DeltaR}
\Delta_{\rm R} = \frac{  \overline{R}_{\rm mem} - \mean{\overline{R}_{\rm mem}|\lambda,z}  }{ \mean{\overline{R}_{\rm mem}|\lambda,z}  }.
\end{equation} 
Here $\mean{\overline{R}_{\rm mem}|\lambda,z}$ is the mean $\overline{R}_{\rm mem}$ value at a particular $\lambda$ and $z$ bin, estimated by fitting a spline to the average value of $\overline{R}_{\rm mem}$ in ten bins of $\lambda$ and five bins of $z$.
By construction, negative $\Delta_{\rm R}$ value means the cluster has a more compact member galaxy
distribution than the average cluster at that richness and redshift.

\subsection{Galaxy shape data}
\label{sec:shape}

In this work, we use 3 different galaxy shape measurement methods from 4 catalogs to determine the galaxy
position angle and ellipticity, to investigate systematics in the measured central galaxy alignment
signal. This section includes a description of all of these methods.

\subsubsection{Re-Gaussianization shape measurement}

The first shape measurement method is based on the re-Gaussianization technique \citep{Hirata03},
which not only corrects the effects of the point spread function (PSF) on the observed galaxy shapes
with a standard elliptical Gaussian profile, but also corrects for low-order deviations
from Gaussianity in both the galaxy and PSF profiles.

Two shape catalogs generated using the re-Gaussianization technique are used in this work; the
primary one is based on the SDSS DR8 photometric pipeline, and was presented in R12; however, for
systematics tests we also use the catalog from M05, which was based on the DR4 photometric pipeline. The R12 catalog covers an area of 9432 deg$^2$, with
an average of 1.2 galaxies arcmin$^{-2}$ with shape measurements; the M05 catalog covers an area of
7002 deg$^2$. Both shape catalogs select galaxies down to the extinction-corrected r-band model
magnitude $m_r < 21.8$, and require galaxies to be well resolved compared to the PSF size in both
$r$ and $i$ bands. While there are minor differences in galaxy selection criteria in the catalogs,
the main difference is the version of the SDSS photometric pipeline 
({\tt photo}) that they used. The M05 catalog relies on {\tt photo} v5.4 \citep{Adelman-McCarthy06} while
the R12 is based on {\tt photo} v5.6 \citep{Aihara11}.  The new version of {\tt photo} has
a more sophisticated sky-subtraction algorithm that improves the photometry 
of large galaxies and fainter ones near them. 
By comparing the central galaxy alignment measured using these catalogs, we will estimate the impact of the sky-subtraction quality on the final results.

\subsubsection{Isophotal shape measurement}

Many previous central galaxy alignment studies used the SDSS isophotal position angle to define the
orientation of the BCG \citep{Brainerd05, Yang06,Faltenbacher07, Azzaro07, Wang08, Siverd09,
  Agustsson10, Hao11}. To compare with these studies, we also measure the central galaxy
alignment using the isophotal shape measurement. The SDSS pipeline measures the isophotal position angle of
galaxies at the isophote corresponding to 25 mag arcsec$^{-2}$, which is fairly low surface
brightness and generally encompasses a much larger part of the galaxy light profile than the
centrally-weighted re-Gaussianization shapes.

Isophotal shapes were not released in DR8, so we take the isophotal position angle in $r$ band from 
DR7 (using the previous version of {\tt photo}) to compute central galaxy alignments.

\subsubsection{De Vaucouleurs shape measurement}

Some galaxy alignment studies use the shape measurement from the de Vaucouleurs model fit
\citep{Niederste-Ostholt10,Siverd09,Hao11}, which is a good description of the surface brightness
profile for a typical elliptical galaxy, including most galaxies in 
redMaPPer clusters. Here we
use the de Vaucouleurs fit position angle provided in the SDSS DR7, which fits galaxies through a
two-dimensional fit to a PSF-convolved de Vaucouleurs profile. For more detail about these SDSS
shape measurements, we refer readers to \citet{Stoughton02}.

\subsection{The central-satellite pair sample}\label{subsec:pairdef}

\begin{figure*}
\begin{center}
\includegraphics[width=0.9\textwidth]{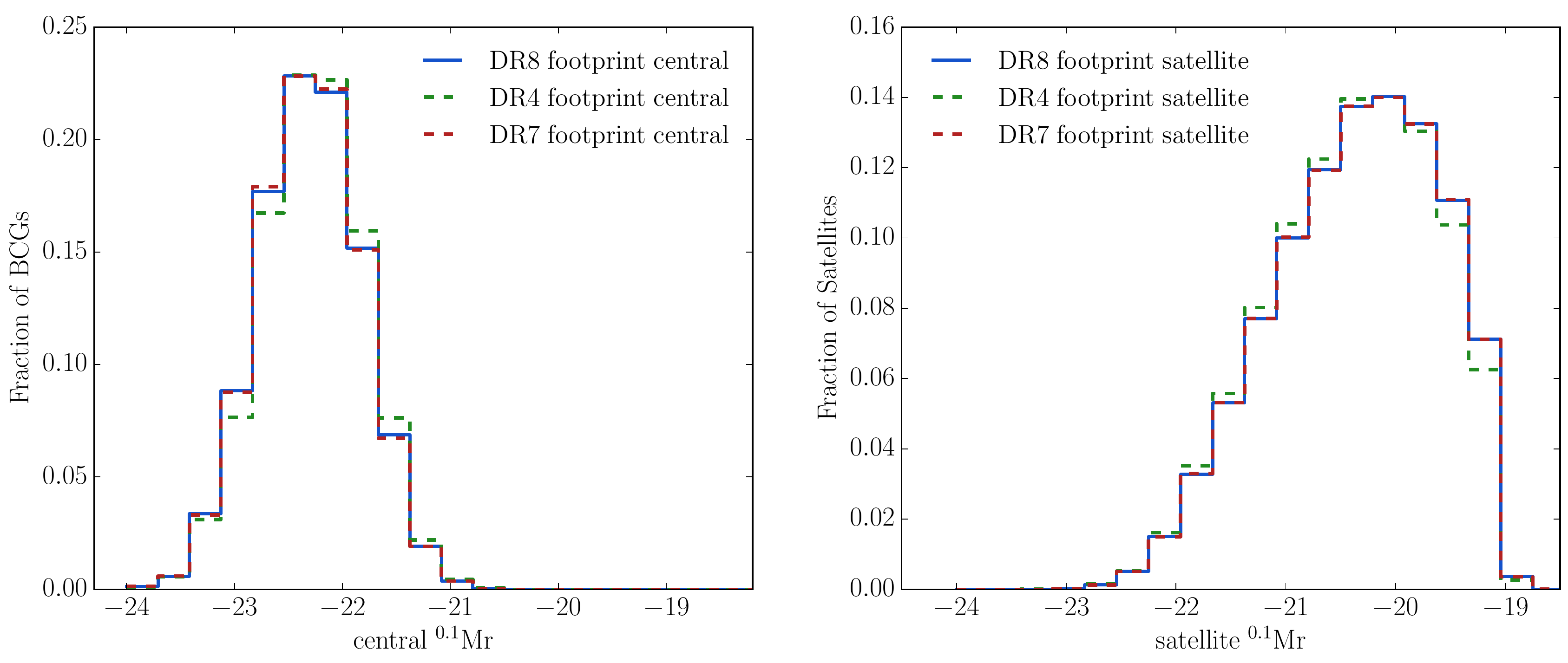}
\caption{Distributions of the central (left panel) and satellite (right panel) $^{0.1}M_r$ for the three
  sets of central-satellite pairs defined in Sec.~\ref{subsec:pairdef}.} 
\label{fig:Mr01_dist} 
\end{center}
\end{figure*}

We define three samples of central-satellite pairs for our analysis:
\begin{enumerate}
\item After applying the redshift cut and requiring that central galaxies have shape measurements in the R12 catalog, we have 8237 centrals with DR8 re-Gaussianization shape measurement,
and 94817 satellites with $p\rm_{mem} \geq 0.8$ in our parent sample.  This parent sample is used
for the majority of our analysis, while the other subsamples are used primarily for systematics tests.

\item To investigate the effect of the sky-subtraction technique on the measured central galaxy alignment signal, we match our parent centrals with the M05 catalog, and construct another sub-sample of centrals that
have re-Gaussianization shape measurement based on both DR4 and DR8 photometry. This subsample has 
 4316 centrals and 46370 central-satellite pairs within the DR4 footprint.
 
\item To compare the degree of central galaxy alignment signal using different shape measurement methods, 
another subsample of central-satellite pairs is constructed. If we require centrals to have a DR8
re-Gaussianization shape, along with both isophotal and de Vaucouleurs shape measurements from DR7,
we have 7488
centrals with 86350 satellites within the DR7 footprint.
\end{enumerate}

Fig.~\ref{fig:Mr01_dist} shows the distributions of the $r$-band absolute magnitude, $^{0.1}M_r$, of
the centrals (left panel) and satellites (right panel) in these three sets of central-satellite pairs.  Both
the central and satellite $^{0.1}M_r$ distributions for the subsample in the DR7 footprint (iii)
are almost the same as for the parent DR8 sample (i), while there are slight shifts
for the subsamples in DR4 footprint (ii).

When measuring the central alignment angle, we only require satellite positions and central shape
measurements. However, in the linear regression analysis, we
require all galaxies to have well-defined physical parameters such as ellipticity, color, effective
radius\dots; these requirements eliminate some satellites, mainly due to the 
requirement of an ellipticity measurement. Table~\ref{tb:sample_LR} summarizes the three sets of
central-satellite pairs defined in this section, and also records the actual
number of central-satellite pairs used when doing linear regression analysis.

\begin{figure}
\begin{center}
\includegraphics[width=0.43\textwidth]{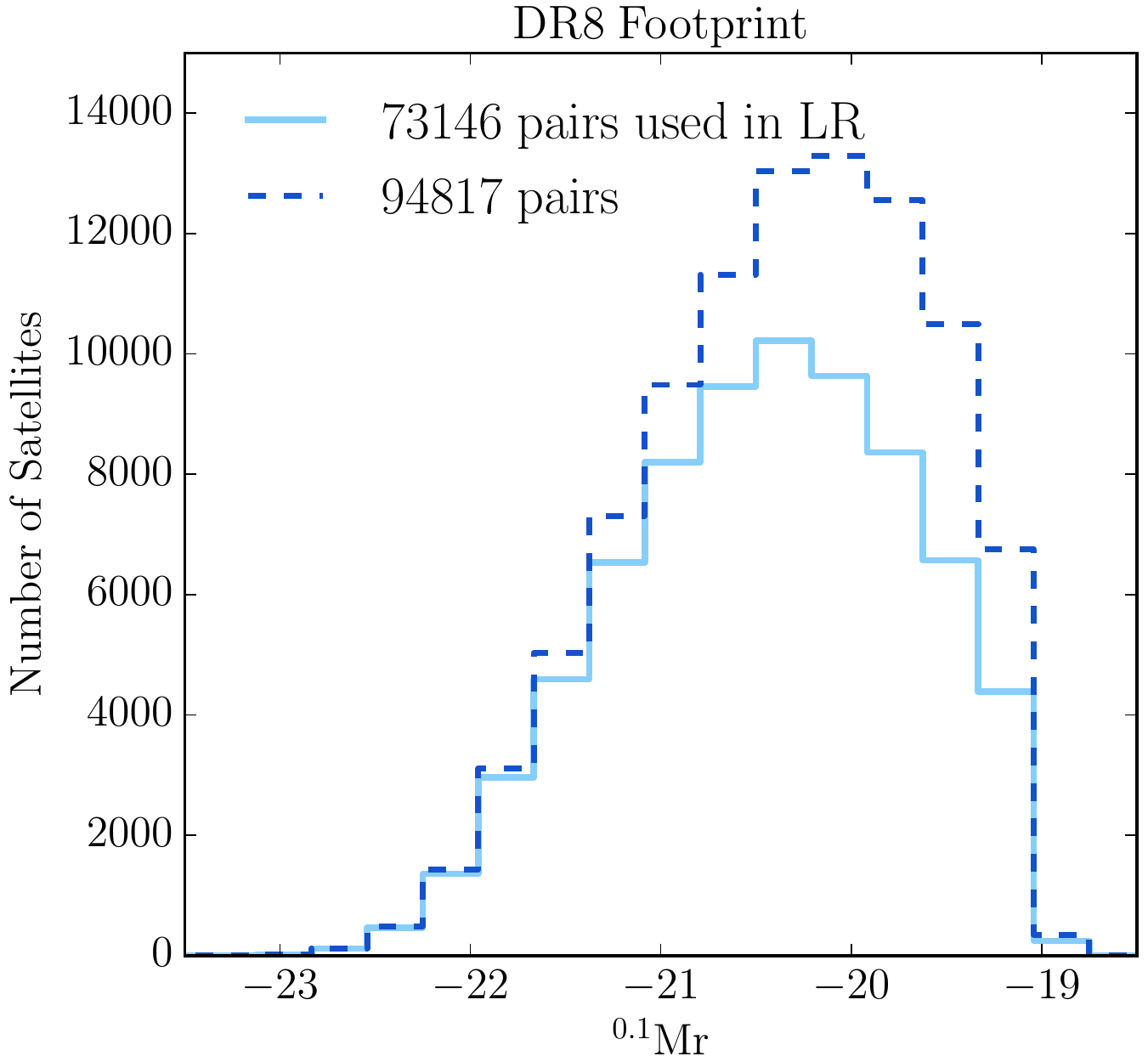}
\caption{Distributions of the $^{0.1}M_r$ for satellites in the DR8 footprint. The dark blue dashed
  line indicates the total 94817 satellites, while the light blue line shows the selected subsample when doing linear regression analysis.}
\label{fig:LR_Mr01_dist} 
\end{center}
\end{figure}

In Fig.~\ref{fig:LR_Mr01_dist}, we compare the absolute magnitude distributions of the satellite
subsamples actually used in the linear regression analysis to that of the original set of satellites from
which they were drawn. The selected satellites used in linear regression are biased to brighter
magnitudes, since we rely on good quality photometry (higher $S/N$ and/or more resolved light
profile) to measure shapes. For the reason, 
the derived significance levels for potential predictors that could possibly affect the degree of
central galaxy alignment in this work are lower limits, especially for predictors that strongly correlate with satellite brightness. 
If the effect of a predictor on central galaxy alignment is strong enough, then even if
some faint satellites are excluded when doing linear regression, we could still select the predictor
out as a featured predictor.

\begin{table}
\caption{Numbers of clusters and central-satellite pairs used in this work. The first three rows are the  the three subsamples we used for the overall measurement of the central galaxy alignment angle defined in Sec.~\ref{subsec:pairdef}. The last row is the subsamples used when doing linear regression analysis.}
\begin{center}
\begin{tabular}{lcccc}
\hline 

Sample                                   	&   N$_{\rm cluster}$ &  N$_{\rm pair}$       \\ \hline  \hline
DR8 Footprint Sample	&    8237    & 94817   \\   \hline

DR4 Footprint Sample	&    4316    & 46370    \\ \hline

DR7 Footprint Sample	&    7488  &   86350     \\ \hline

Linear Regression Sample	&    8233  &   73146     \\

\label{tb:sample_LR}
\end{tabular}
\end{center}
\end{table}


\section{Linear regression analysis}
\label{sec:LR}

Regression is one of the most commonly used methods to study dependence. 
It is used to find optimal values of the free parameters in a specified function
$Y = f(\bold{X}) + \epsilon$. 
Here $Y$ is the response variable, which quantifies the physical effect one wants to study, 
$\bold{X} = (X_1, ... X_i, ... X_N)$ is a set of potential predictors that may affect the behavior of $Y$,
and $\epsilon$ represents random observational error,
usually assumed to be drawn from a normal distribution. 
For the central galaxy alignment effect, as there is no a priori-known functional form relating
$\bold{X}$ to $Y$, we apply {\em multiple linear regression}, which allows one to at 
least determine if the central galaxy alignment depends on $\bold{X}$  
to first order.

The multiple linear regression model we apply is
\begin{equation}\label{eq:LRfun}
Y = f(\bold{X}) = \beta_0+\beta_1 X_1+ \ldots+\beta_i X_i+ \ldots + \beta_N X_N,
\end{equation}
where the intercept $\beta_0$ and the slopes $\beta_i$ are the unknown regression
coefficients to be estimated via least squares.
For each regression coefficient $\beta_i$, we perform the two-sided $t$-value and $p$-value tests
for the dependence of $Y$ on the $X_i$.  These are tests of the hypothesis that $\beta_i = 0$ against the
alternative hypothesis that $\beta_i \ne 0$.  The $t$-value is the ratio of $\beta_i$ to its
standard error, which can be positive or negative depending on the sign of $\beta_i$. A larger $|t|$
indicates a more significant statement that $\beta_i \ne 0$, which means it is more
likely that there is a relationship between $Y$ and $X_i$.  
Statistically, the $t$-value and $p$-value are inextricably linked. 
Under the assumption of normally distributed errors,
a $p$-value of 0.05 corresponds to a 95\% confidence that
$\beta_i$ is not equal to zero. 
Thus we select out a regressor $X_i$ as a featured predictor if
its $p$-value $< 0.05$ \citep[e.g.,][]{Weisberg13}.

There are reasons not to use our predictors $P_i$ defined in Sec.~\ref{subsection:definitions}
(cluster ellipticity, central galaxy dominance, etc.) directly as regressors $X_i$.  Since there is a large
variation in the range of each predictor, if we simply regress by $Y =
\beta_0+\sum_i \beta_i\p P_i$, the fitted magnitude of
$\beta_i\p$ would depend on that range, i.e., for a given level of
correlation between the parameters, $\beta_i\p$ would be small if its $P_i$ tends to be large. 
To make our results more directly illustrate how the relative change of a physical parameter affects the
value of $Y$, throughout we normalize our predictors $P_i$ to obtain regressors as follows:
\begin{equation}\label{eq:normalization}
X_i = \frac{P_i - \mean{P}_i}{\sigma_{P_i}}.
\end{equation}
Here $\sigma_{P_i}$ is the sample standard deviation of the predictor $P_i$, reflecting the width of the intrinsic distribution and measurement error.
We will use the term
``predictor'' to correspond to the original variables, and  ``regressor'' to refer to
variables that are transformed as in Eq.~\eqref{eq:normalization}. 
We note, however, that using the normalized predictors as our regressors does not affect the result
of hypothesis tests to select featured predictors.

In this work, we will build two multiple linear regression models with two different response variables, 
and use a total of 16 potential predictors to analyze the central galaxy alignment and the angular segregation of satellites. 
Details on the definitions and measurements of these physical parameters were presented in Sec.~\ref{subsection:definitions}.

\subsection{Response variables}

The two response variables used to quantify the level of central alignment are: 1) the position angle
difference between the central galaxy and its host cluster 
\begin{equation}\label{eq:PA_difference}
\Delta \eta = |P.A._{\rm cen} - P.A._{\rm cluster}|\rm ,
\end{equation}
and 2) the central galaxy alignment angle for each central-satellite pair, $\theta_{\rm cen}$.  We use these in different ways as described below.

$\Delta \eta$ lies in the range $[0^{\circ}, 90^{\circ}]$, where $0^{\circ}$ indicates that the central galaxy is perfectly aligned with the shape of the projected member galaxy distribution of the cluster.
That distribution is believed to trace the underlying DM halo shape with some scatter
\citep{Evans09, Oguri10}. The quantity $\Delta \eta$ is thus an observable proxy for the level of
central galaxy alignment with its DM halo. We will regress it onto central galaxy- and
cluster-related predictors to identify what central galaxy properties and/or cluster properties most
strongly predict the alignments of central galaxies with their satellite galaxy distributions.

The definition of $\theta_{\rm cen}$ is illustrated in the left panel of Fig.~\ref{fig:illustration}.
It is a direct observable reflecting a satellite's angular position with respect to the major axis direction of 
it's central galaxy.
With each satellite galaxy having its corresponding $\theta_{\rm cen}$ as the response variable, 
we will regress it onto individual satellite quantities to understand what kind of satellites are more
preferentially located along the major axis of the central galaxy. 

\subsection{Potential predictors}
\label{subsection:predictors}

We classify the 16 predictors into three categories: central-related, cluster-related and satellite-related quantities. 
Table \ref{tb:predictors} lists these predictors under each category.  

\begin{figure}
\begin{center}
\includegraphics[width=0.9\columnwidth]{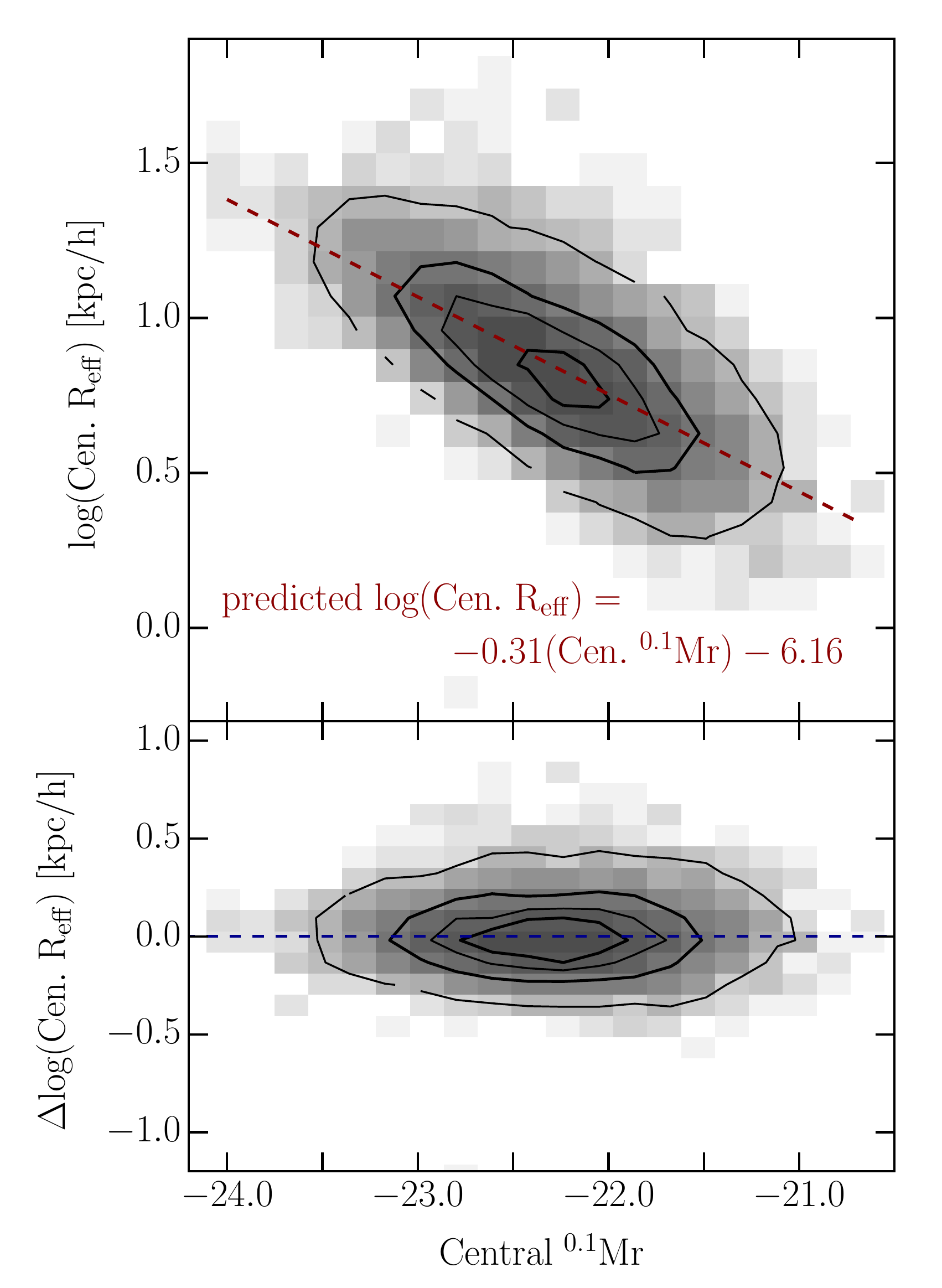}
\caption{The top panel shows the contour plot of log(central R$\rm_{eff}$) versus central $^{0.1}M_r$. The
  red dash line shows the  least-squares
  fitting of a linear relationship between  log(central
  R$\rm_{eff}$) and central $^{0.1}M_r$, with the equation of the best fitted line shown on the plot. The bottom panel plots the residuals versus central $^{0.1}M_r$.}
\label{fig:Reff_BCG}
\end{center}
\end{figure}

\textbf{Central Galaxy Quantities:} 
We use six central galaxy related physical parameters: central galaxy dominance, $^{0.1}M_r$, $^{0.1}M_g$-$^{0.1}M_r$ color, ellipticity, effective radius, and central probability. 
Since there is a tight correlation between the size and luminosity of galaxies
\citep[e.g.,][]{Bernardi14}, in order to investigate the effect of galaxy size on central galaxy alignments, we
use the offsets in galaxy size from the fitted size-magnitude relation, $\Delta$log(cental
$R\rm_{eff}$) $\equiv$ measured log(cental $R\rm_{eff}$) $-$ predicted log(cental $R\rm_{eff}$), as our
predictor when doing linear regression. The top panel of Fig.~\ref{fig:Reff_BCG} shows the log(central R$\rm_{eff}$)--central $^{0.1}M_r$ correlations for the DR8 central galaxies.
In the bottom panel, we present the log(central R$\rm_{eff}$) residuals from the fitted log(central R$\rm_{eff}$)--central $^{0.1}M_r$ relation, as a function of central $^{0.1}M_r$. 

\textbf{Cluster Quantities:} We have four cluster-related physical parameters: log(richness), redshift, 
cluster ellipticity and cluster member concentration $\Delta_{\rm R}$. 

\begin{figure}
\begin{center}
\includegraphics[width=0.9\columnwidth]{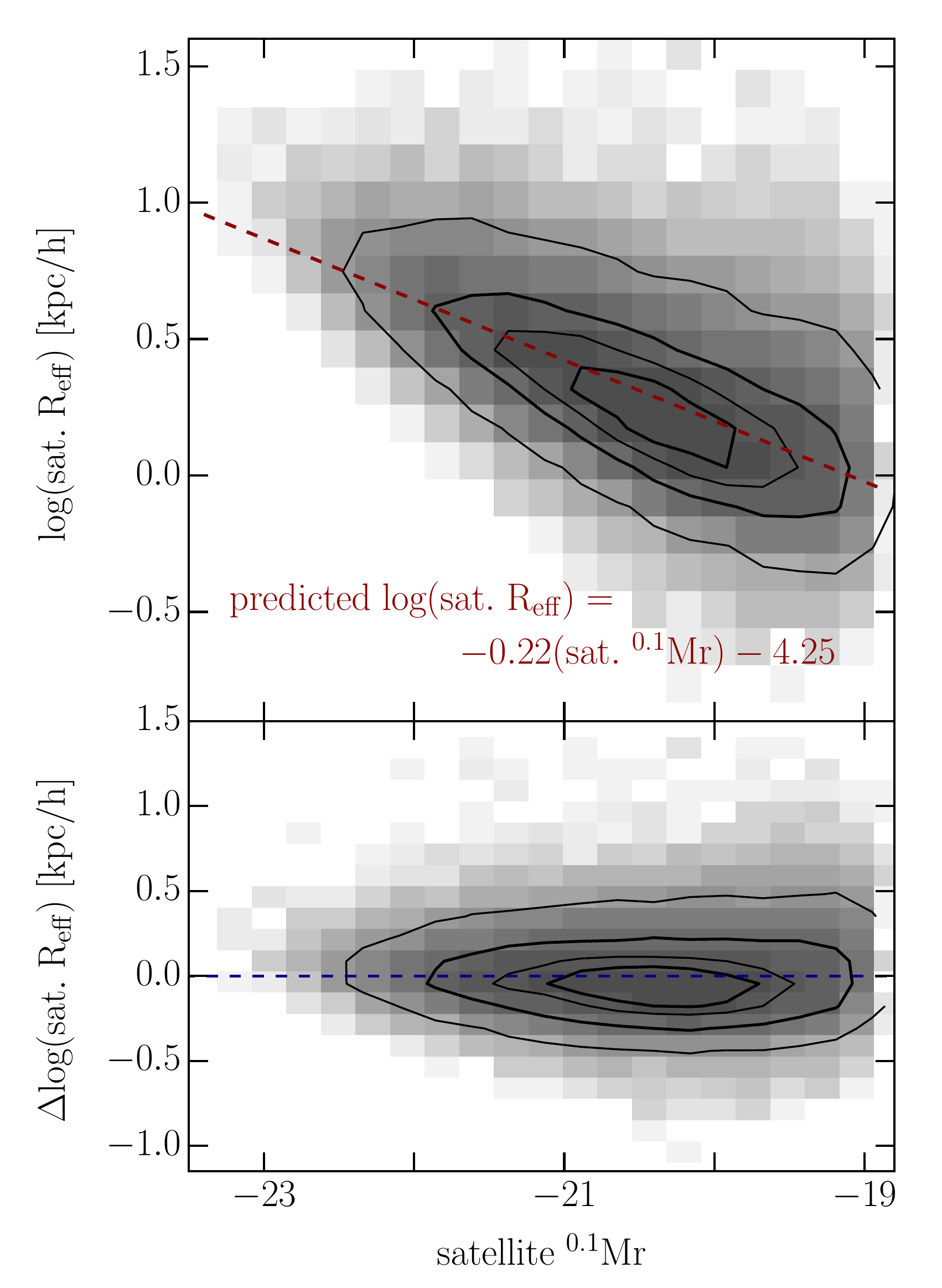}
\caption{Similar to Fig.~\ref{fig:Reff_BCG} but for cluster member galaxies instead of centrals.}
\label{fig:Reff_sat}
\end{center}
\end{figure}

\textbf{Satellite Quantities:} The six satellite-related quantities are the cluster-centric distance
of each satellite normalized by its host $R_{\rm 200m}$, $^{0.1}M_r$, $^{0.1}M_g$-$^{0.1}M_r$ color,
ellipticity, effective radius, and the satellite alignment angle $\phi_{\rm sat}$. As for the central galaxies, we use the residual effective radius, $\Delta$log($R\rm_{eff}$), for satellites as our
physical parameter instead of the directly measured $R\rm_{eff}$, in attempt to eliminate the
contribution of luminosity on size. We fit the log(R$\rm_{eff}$)--$^{0.1}M_r$ correlations for the
the 73146 satellites as shown in the top panel of Fig.~\ref{fig:Reff_sat} first, and then use
$\Delta$log($R\rm_{eff}$) $\equiv$ measured log($R\rm_{eff}$) $-$ predicted log($R\rm_{eff}$) as our new
size predictor. The bottom panel of  Fig.~\ref{fig:Reff_sat} shows the $\Delta$log($R\rm_{eff}$) as
function of $^{0.1}M_r$.  

\begin{table*}
\caption{The 16 potential predictors used to study the central galaxy alignment effect in this
  work.}
\begin{center}
\begin{tabular}{lclcl}
\hline 

\textbf{Central Galaxy Quantities}	&  &  \textbf{Cluster Quantities}	&  & \textbf{Satellite Quantities}    \\ \hline  \hline
         central galaxy dominance		&  &  log(richness)			&  &  log($r/R_{\rm 200m}$)	\\
         central $^{0.1}M_r$			&  &  redshift				&  &  satellite $^{0.1}M_r$		\\
	central $^{0.1}M_g-^{0.1}M_r$ color &  &  cluster ellipticity		&  &  satellite $^{0.1}M_g-^{0.1}M_r$ color \\
	central ellipticity 			&  &  cluster member concentration $\Delta_{\rm R}$  &  &  satellite ellipticity \\
	$\Delta$log(central $R_{\rm eff}$) &  &  					&  &  $\Delta$log(satellite $R_{\rm eff}$) \\
	$P_{\rm cen}$				&  &  					&  &  $\phi_{\rm sat}$  \\

\label{tb:predictors}
\end{tabular}
\end{center}
\end{table*}

\subsection{Variable selection}\label{subsec:selection}

The goal of variable selection \citep[e.g.,][]{burnham2003, James13} is to identify the subset of predictors that are
important within a large pool of potential predictors. 
There are many statistical methods for subset selection; we adopt 
the ``Forward-Stepwise Selection'' approach \citep[See e.g., Sec. 3.3 of][]{friedman2001elements}. Beginning with a model containing no predictor, forward-stepwise
selection involves fitting $N$ models for the $N$ predictors separately: $Y = \beta_0 +
\beta_i X_i$, and selects the regressor $X_p$ with the most significant hypothesis test on $\beta_p
\ne 0$, i.e., greatest absolute $t$-value or smallest $p$-value. In the second cycle, $N-1$ models
for the remaining $N-1$ predictors are fit via $Y = \beta_0 + \beta_p X_p + \beta_i X_i$, where $i
\ne p$, and again we select the most significant regressor $X_q$.  At each stage, one predictor is
selected to add to the model until the remaining regressors have $p$-value $> 0.05$, 
which is a common stoping choice in many statistical packages.
The forward stepwise algorithm therefore considers at most $N + (N-1) + \ldots + 1 = N(N+1)/2$ models
in the extreme case when all $N$ regressors have $p$-value $< 0.05$.  
We then fit a model using least squares on the reduced set of variables, and determine the final $t$- and $p$-values of the selected featured predictors. 

To ensure the robustness of our variable selection scheme,
we compare our variable selection result with another variable 
selection method -- ``Best-Subset Selection'' -- which considers all 2$^N$ 
possible combinations of models from the $N$
predictors, and selects the best one based on a model-selection criterion, such as Mallow's $C_p$ \citep{mallows1973}, Akaike information criterion (AIC, \citealt{akaike1998}), Bayesian information criterion (BIC, \citealt{schwarz1978}), or adjusted $R^2$.
(By contrast, forward-stepwise selection is a so-called
greedy algorithm -- at each step, it selects only that one regressor that best
improves the overall fit -- and thus it can fail to uncover the optimal 
model. However, it does have the virtue of computational efficiency.)
Different section criterion places different penalty on the complexity of the model. BIC penalizes heavier on models with more variables and hence tends to select smaller number of predictors, while adjusted $R^2$ puts less penalty thus results in selecting more predictors.
In this work we use $p$-value $< 0.05$ as the criteria to pin down the total number of predictors in the forward-stepwise selection process, and this result agrees with that from best-subset selection under Mallow's $C_p$ and AIC, validating our use of
forward-stepwise selection with our dataset.

Throughout this work, we use the statistical package {\tt StatsModels} in Python to do forward
stepwise selection, and use the {\tt leaps} package in R to perform best-subset selection.

In this work, we attempt to address two main questions: 1) What central galaxy and cluster properties are the strongest
predictors of the strength of central galaxy alignments? 2) What kinds of satellites are more likely
to lie along the major axis direction of their host central galaxy?
To address the first question, we regress $\Delta \eta$ against the central- and cluster-related
quantities, and use forward-stepwise selection to pick featured predictors. Once we have a good
model in terms of central- and cluster-related quantities, we move to the second question by using
$\theta_{\rm cen}$ as a response for each central-satellite pair, and regress $\theta_{\rm cen}$ against
the individual satellite quantities.   
To isolate the effects of satellite properties, we must properly account for the overall effect from
their host central galaxies and clusters.  Thus, we start with a model containing the selected central and cluster
predictors from the previous stage, and use the forward-stepwise procedure to see whether (with the
presence of these central and cluster quantities) there are also satellite quantities that are
significant enough to be selected  as featured predictors. 


\section{Results}
\label{sec:result}

In this section, we report the results of an analysis of central galaxy alignments, including  our linear regression analysis.

\subsection{Overall signal}

\subsubsection{Distribution of $\Delta \eta$}
\begin{figure*}
\begin{center}
\includegraphics[trim=0.15cm 0cm 1.0cm 0cm,width=1.0\textwidth]{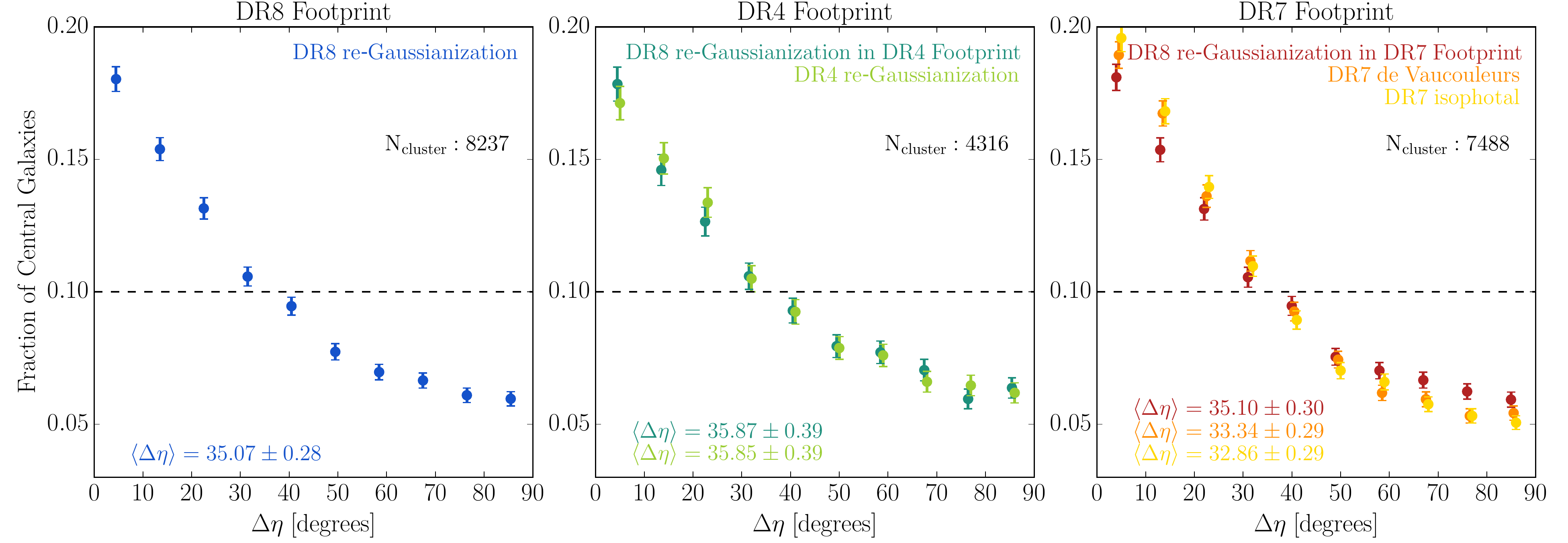}
\caption{Distributions of the position angle difference between central galaxy and cluster, $\Delta \eta=
  |P.A._{\rm cen} - P.A._{\rm cluster}|$. The left panel shows the $\Delta \eta$ distribution of the
  parent 8237 centrals measured by the re-Gaussianization method in DR8. The middle panel plots the
  $\Delta \eta$ distributions of the 4316 centrals which are both measured by the re-Gaussianization
  method in the DR4 footprint, with the light (dark) green dots representing measurements based on
  the DR4 (DR8) photometry. The right panel shows the $\Delta \eta$ distributions of the 7488 centrals
  in DR7 footprint measured by re-Gaussianization (red dots), de Vaucouleurs (orange dots), and isophote
  (yellow dots) methods respectively. Points are slightly shifted horizontally for clarity. Error
  bars indicate the standard error of the mean. 
The horizontal black dashed line indicates the prediction for randomly-oriented central galaxies. 
The mean position angle difference, $\mean{\Delta \eta}$, is shown in the bottom left
corner of each panel.   
}
\label{fig:Delta_eta_dist} 
\end{center}
\end{figure*}

We begin with a basic analysis of the properties of central galaxy
alignments. 
Fig.~\ref{fig:Delta_eta_dist} shows the distributions of the position angle difference between the central galaxy
and cluster shapes, $\Delta \eta$, for our three cluster samples 
tabulated in Table~\ref{tb:sample_LR}, defined for the purpose of investigating systemics in various shape measurement techniques.
The distributions show a highly significant degree of central galaxy alignment with cluster orientations. 
The bottom left corner of each panel shows the average $\Delta \eta$ value, $\mean{\Delta \eta}$, for
each sample; they are all $\mean{\Delta \eta} < 45^{\circ}$ at high significance.
Hence, if indeed the satellite galaxy distributions trace the dark matter halo shapes, then centrals
also tend to align with their underlying halos.  

The left panel of Fig.~\ref{fig:Delta_eta_dist} is the $\Delta \eta$ distribution of the 8237 centrals
measured by the re-Gaussianization method in DR8. The average $\Delta \eta$ for parent dataset is
$\mean{\Delta \eta}=35.07^\circ\pm0.28^\circ$; this represents our primary result in this section,
with the remaining results serving as systematics tests.

To compare the effect of sky-subtraction algorithm on the signal, we use the sample of 4316 centrals
in the DR4 footprint with re-Gaussianization shape measurements using both DR8 and DR4
photometry. The middle panel of Fig.~\ref{fig:Delta_eta_dist} shows the $\Delta \eta$ distributions
of this sample, with the light (dark) green dots indicating measurements based on the DR4 (DR8)
photometry. 
Within the error bars, the two $\Delta \eta$ distributions and their mean $\mean{\Delta \eta}$
values are consistent with each other. 
We therefore conclude that for the re-Gaussianization shapes, the effect of sky-subtraction does not
substantially influence the overall distribution of $\Delta \eta$. However, this conclusion may not
be applicable for other shape measurement methods that trace different regions on the surface
brightness profile of galaxies.  Re-Gaussianization shapes are weighted more toward the inner part
of the light profile, which is less sensitive to sky subtraction errors, while isophotal shapes are
more sensitive to the outer part and could have more systematics due to sky subtraction.  However,
due to the lack of isophotal shapes in DR8, we cannot test this effect by comparing different data reductions.

To investigate the effect of shape measurement methods on the detection of central galaxy alignments, we use
the sample of 7488 BCGs that have re-Gaussianization, de Vaucouleurs, and isophotal shape
measurements in DR7 footprint. The right panel of Fig.~\ref{fig:Delta_eta_dist} shows the $\Delta \eta$
distributions of these samples, with the red, orange, and yellow dots representing shape
measurements based on the re-Gaussianization method, de Vaucouleurs fit, and isophotal fit, respectively. 
Within the error bars, the second and third distributions agree, while the $\Delta \eta$
distribution measured using the re-Gaussianization method differs systematically.  
The value of $\mean{\Delta \eta}$ using the re-Gaussianization method ($35.10^\circ\pm0.30^\circ$)
is significantly larger than that calculated by de Vaucouleurs ($33.34^\circ\pm0.29^\circ$) and
isophotal ($32.86^\circ\pm0.29^\circ$) shape measurements. 
This could be due to a systematic or caused by a true physical effect. We will discuss in detail in Sec.~\ref{sec:shape-align}.

\begin{figure}
\begin{center}
\includegraphics[width=0.5\textwidth]{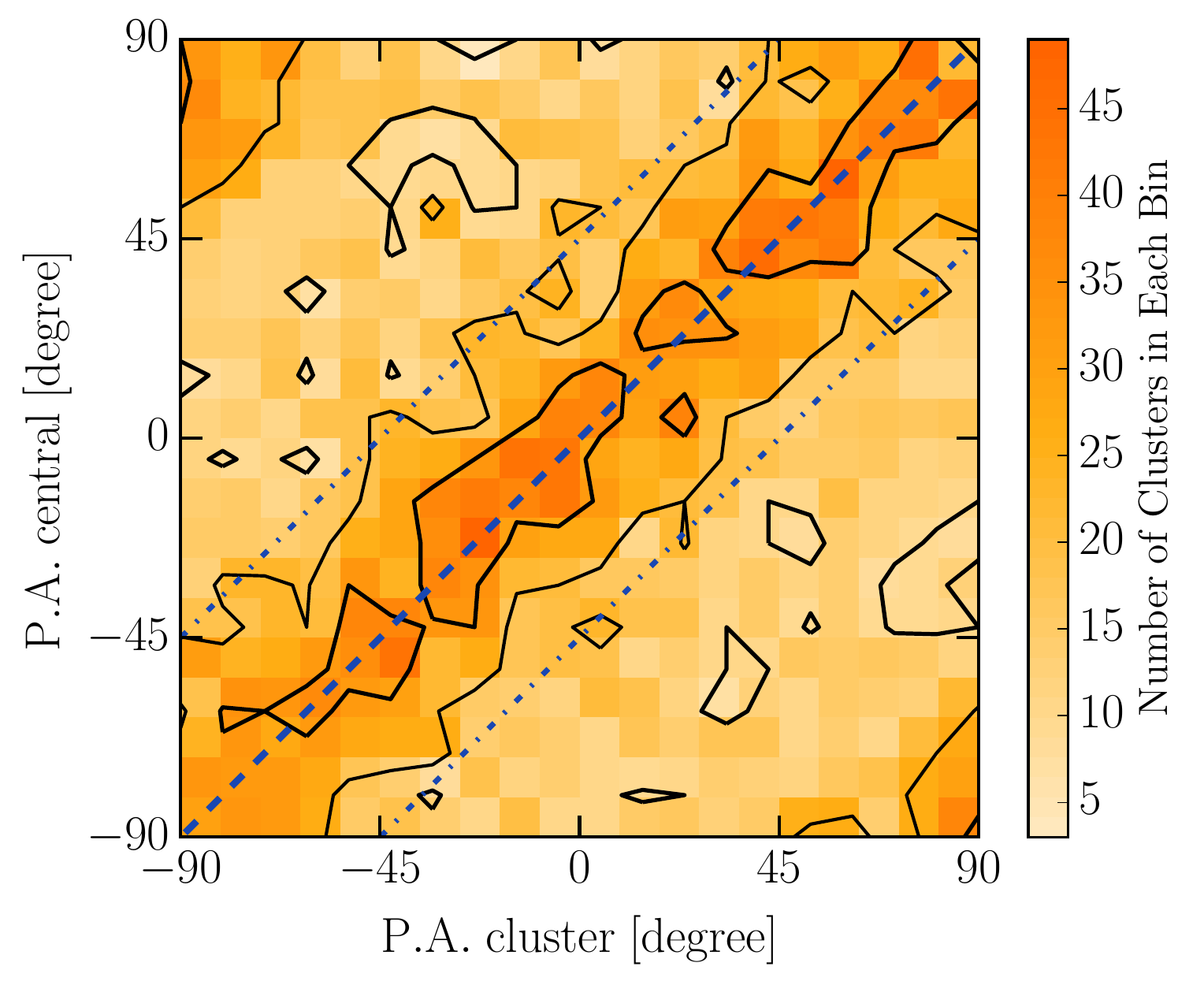}
\caption{Comparison between the position angles of cluster and central galaxy. The blue dashed line indicates
  the case where the position angle of the central is the same as that of its cluster. 
  The two blue dot-dashed lines delineate a region where the P.A. differences between the cluster and central galaxy are less than 45$^\circ$.}
\label{fig:cpPA}
\end{center}
\end{figure}

As further illustration of the alignment between the central and shape of member galaxy distribution, in
Fig.~\ref{fig:cpPA} we compare the cluster and central galaxy position angles. With the overall distribution
peaking around the symmetric axis of the figure, we observe the preference for centrals pointing toward
the orientation directions of clusters. 

\subsubsection{Distribution of $\theta\rm_{cen}$}

\begin{figure*}
\begin{center}
\includegraphics[trim=0.15cm 0cm 1.0cm 0cm,width=1.0\textwidth]{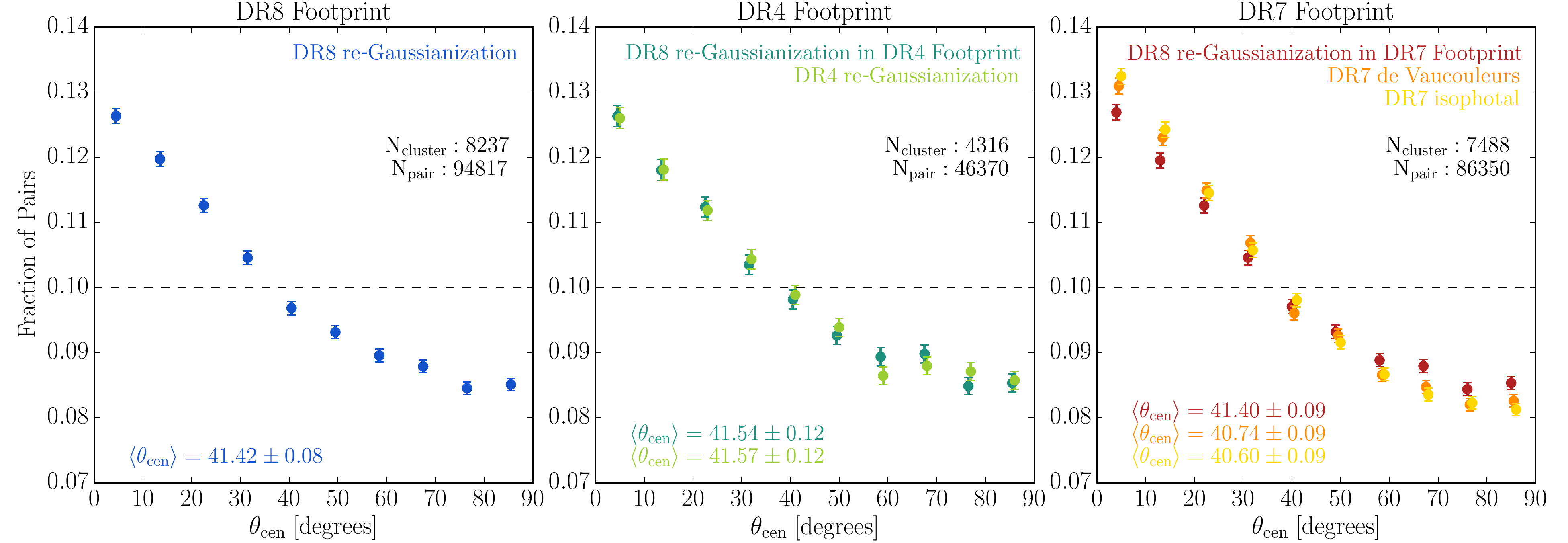}
\caption{Distributions of the central galaxy alignment angle. The left panel shows the $\theta\rm_{cen}$
  distribution of the 94817 central-satellite pairs measured by the re-Gaussianization method in
  DR8. The middle panel shows the $\theta\rm_{cen}$ distributions of the 46370 central-satellite pairs
  measured by re-Gaussianization in both DR8 photometry (dark green) and DR4 photometry (light
  green), within the DR4 footprint.  The right panel shows the $\theta\rm_{cen}$ distributions of
  the 86350 central-satellite pairs in DR7 that are measured by re-Gaussianization (red dots), de Vaucouleurs (orange dots), and isophote (yellow dots) methods respectively. Points are slightly shifted horizontally for clarity. Error bars are represented by the standard error of the mean. 
The horizontal black dash line indicates the case if satellites were isotropically distributed
around centrals. The mean central galaxy alignment angle, $\mean{\theta\rm_{cen}}$, is shown at the bottom left corner of each panel. 
}
\label{fig:BCG_align_ang_dist} 
 
\end{center}
\end{figure*}

Fig.~\ref{fig:BCG_align_ang_dist} shows the distributions of the central galaxy alignment angle,
$\theta\rm_{cen}$, for our three sets of central-satellite pairs.  
The preferential alignment of satellites along the central galaxy major axis is quantified in the average central galaxy
alignment angle, $\mean{\theta\rm_{cen}}$, in the bottom left corner of each panel. 
The alignment signal looks less dramatic as revealed in $\mean{\theta\rm_{cen}}$ value compared with $\mean{\Delta \eta}$ shown in Fig.~\ref{fig:Delta_eta_dist}. 
This is because $\theta\rm_{cen}$ records the individual location of each satellite with respect to
its central galaxy major axis; these tend to be more randomized than simply considering the overall satellite distribution as a whole.  

The left panel of Fig.~\ref{fig:BCG_align_ang_dist} shows the $\theta\rm_{cen}$ distribution of the
94817 central-satellite pairs measured by the re-Gaussianization method in DR8. The average central galaxy
alignment angle for this dataset is $\mean{\theta\rm_{cen}}=41.42^\circ\pm0.08^\circ$.

The middle panel shows the $\theta\rm_{cen}$ distributions for the 46370 central-satellite pairs
in the DR4 footprint, using DR8 (dark green) and DR4 (light green) photometry, constructed to compare the effect of sky-subtraction technique on the
measurement of $\theta\rm_{cen}$. Within the error bars, the
two $\theta\rm_{cen}$ distributions and the derived $\mean{\theta\rm_{cen}}$ values are consistent
with each other.  As for $\Delta\eta$, we conclude that for the re-Gaussianization shapes, use of
different SDSS photometry pipelines does not influence the results, but caution that this argument may not hold for other shape measurement methods.

To fairly compare our $\theta\rm_{cen}$ measurement with studies based on different shape
measurement methods, the right panel of Fig.~\ref{fig:BCG_align_ang_dist} shows $\theta\rm_{cen}$
measured via re-Gaussianization (red dots), de Vaucouleurs (orange dots), and isophote (yellow dots)
shape measurements of the 86350 central-satellite pairs in the DR7 footprint. 
Within the error bars, the histograms of $\theta\rm_{cen}$ using de Vaucouleurs and isophotal shapes
are consistent with each other, but that for re-Gaussianization method is systematically different,
resulting in a systematically higher  $\mean{\theta\rm_{cen}}$ ($41.40^\circ\pm0.09^\circ$). We will discuss the measurement difference in Sec.~\ref{sec:shape-align}.

\subsection{Linear regression: central galaxy alignments with satellite distributions}
\label{subsection:LR BCG-DM}

To investigate the alignment of the central galaxy with its dark matter Halo, we use our observational proxy (the difference in central and cluster position angles, $\Delta \eta$) as the response variable, and apply
forward-stepwise selection described in Sec.~\ref{subsec:selection} to select featured predictors among the central- and cluster-related
quantities. The list of predictors is defined in Sec.~\ref{subsection:predictors}, with the
observational method for determining them in Sec.~\ref{subsection:definitions}. 

\begin{figure*}
\begin{center}
\includegraphics[trim=0.5cm 0cm 1cm 0cm,width=1\textwidth]{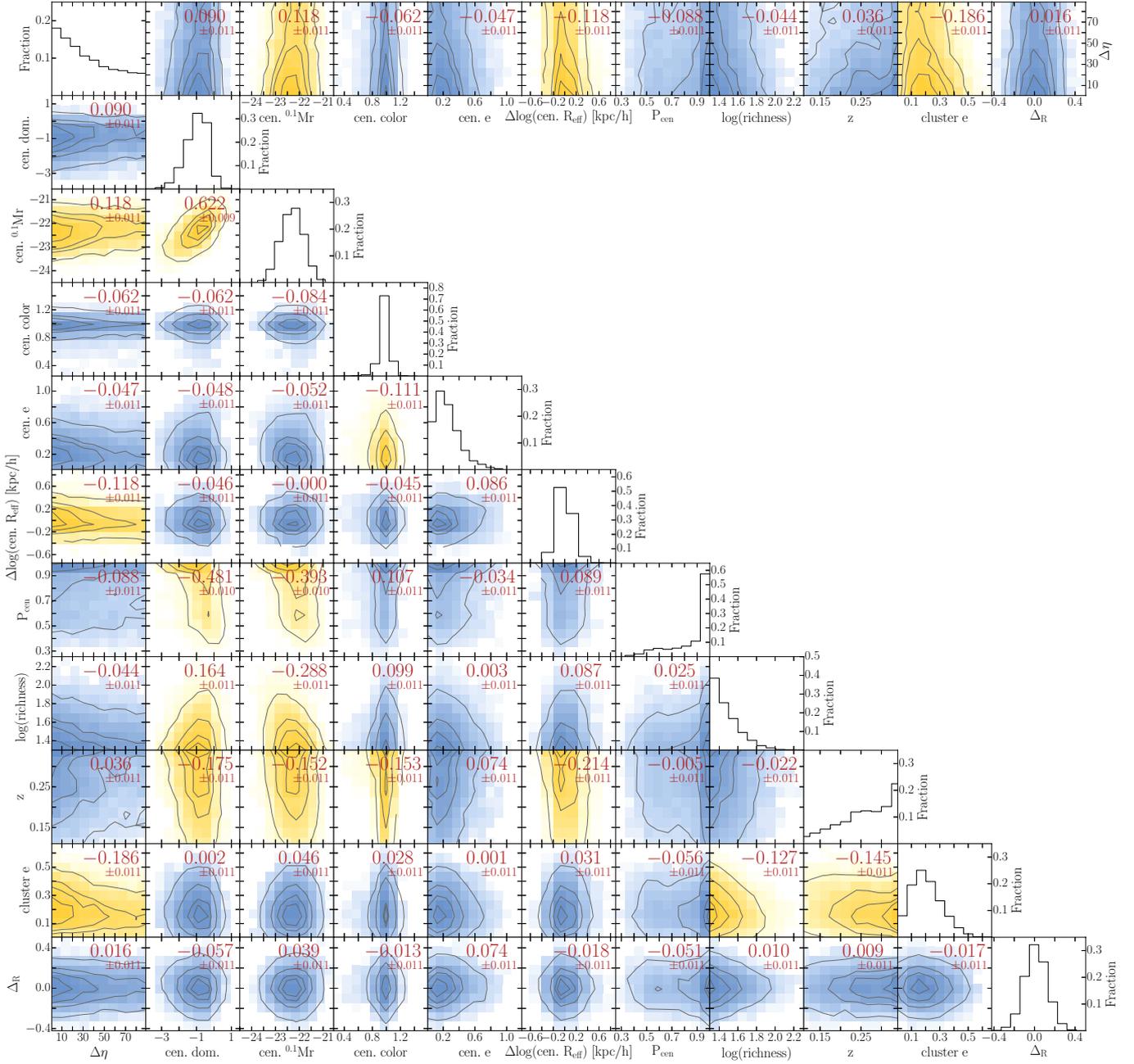}
\caption{Scatterplot matrix of the position angle difference between central and cluster shapes, $\Delta
  \eta$, with the ten central- and cluster-related predictors. The correlation coefficient between each
  pair of parameters is noted on the plot. We highlight scatterplots with correlations that are significant at $>10\sigma$ in yellow. 
  The gray contour levels indicate 20\%, 40\%, 70\%, and 95\% number of clusters of our data.
}
\label{fig:dr8_DM_8233}
\end{center}
\end{figure*}

Fig.~\ref{fig:dr8_DM_8233} displays the scatterplot matrix between $\Delta \eta$ and all of the central-
and cluster-related predictors based on the re-Gaussianization shape measurement. The diagonal panels are
histograms of physical parameters, and the other panels are scatterplots between pairs of
parameters, with the corresponding correlation coefficient noted on each plot. 

Several important results are evident in this scatterplot matrix. 
First, the top row summarizes how $\Delta \eta$ is related to all ten predictors. The sign of the correlation coefficient reveals the direction of the relationship between $\Delta \eta$ and the regressor
, while the magnitude of the correlation coefficient indicates the strength of this dependence. 
The overall impression is that $\Delta \eta$ is weakly related to most of the predictors, with a
maximum  correlation coefficient of $\sim -0.2$ with cluster ellipticity. Though the correlations
are weak, we can still judge whether these
dependences are statistically significant given our large sample size.
Second, some of the predictors are highly correlated with each other, such as central galaxy $^{0.1}M_r$, central galaxy dominance and $P_{\rm cen}$. The forward-stepwise selection procedure will help determine whether we
should keep them all as featured predictors; if any are jointly responsible for the same variation in $\Delta \eta$, then we will select just the most representative one among them.

\begin{table}
\caption{Selected featured predictors for the central galaxy-cluster alignment effect based on the 8233 DR8 clusters. 
The first column is the name of the selected predictor. The second column gives the regression
coefficient $\beta$. Columns 3 and 4 provide the $t$- and $p$-values from the significance tests on
the deviation of $\beta$ from zero. Higher $|t|$ or smaller $p$ indicates a higher significance for
$\beta \neq 0$. Columns 5 and 6 are the mean and standard deviation of the corresponding predictor
for the 8233 clusters, necessary in Eq.~\eqref{eq:normalization} to normalize our predictor $P_i$ to
regressor $X_i$. }
\begin{tabular}{lrrlrr}
\hline 

Predictor						& $\beta$	& $t$-value & $p$-value		& mean	& $\sigma$	 \\ \hline

cluster $e$ 					& -4.91	& -17.8	& $8\times10^{-70}$	& 0.21	& 0.11		\\
$\Delta$log(cen.\ R$\rm_{eff}$)		& -2.67	& -9.7	& $6\times10^{-22}$	& 0.00	& 0.15		\\	
cen. $^{0.1}M_r$				& 2.42	& 7.8		& $9\times10^{-15}$	& -22.30	& 0.47		\\
cen. color						& -1.29	& -4.6	& $4\times10^{-5}$	& 0.97	& 0.08		\\									
$P_{\rm cen}$					& -1.21	& -4.0	& $6\times10^{-5}$	& 0.87	& 0.17		\\
cen. $e$$^\dagger$				& -1.03	& -3.7	& 0.0002			& 0.26	& 0.16		\\
log(richness)					& -0.65	& -2.2	& 0.03			& 1.48	& 0.15		\\

\hline
\multicolumn{6}{l}{$^\dagger$The relationship between central galaxy ellipticity and the} \\ 
\multicolumn{6}{l}{central galaxy alignment signal is more complicated. We will} \\
\multicolumn{6}{l}{provide further investigation in Sec.~\ref{sec: BCG e}.} \\

\label{tb:predictor_bcg}
\end{tabular}
\end{table}

After performing forward-stepwise selection, we find that cluster ellipticity is the most dominant
predictor for the central galaxy alignment effect, and almost all of the central-related quantities are selected as
feature predictors except for central galaxy dominance. The linear regression results, including the
statistical significance of each selected predictor, are in Table~\ref{tb:predictor_bcg}, and the estimated best-fitting equation is: 
\begin{equation}\label{eq:LR1}
\begin{split}
\Delta \eta = 35.07 - 4.91\ \frac{{\rm cluster}\ e-0.21}{0.11}  \\
-2.67\ \frac{\Delta {\rm log(cen.}\ R_{\rm eff})-0.00}{0.15} +2.42\ \frac{{\rm cen.}\ ^{0.1}M_r+22.30}{0.47}  \\
-1.29\ \frac{{\rm cen.\ color}-0.97}{0.08}-1.21\ \frac{P_{\rm cen}-0.87}{0.17}\\
 - 1.03\ \frac{{\rm cen.}\ e-0.26}{0.16} - 0.65\ \frac{{\rm log(richness)}-1.48}{0.15}
\end{split}
\end{equation}

Here we note that since the relation between $\Delta \eta$ and these selected predictors is
  not truly linear and has substantial stochasticity, 
we cannot rely on the resulting regression equation to predict the value of $\Delta \eta$ for any
given cluster.  We 
can only use Eq.~\eqref{eq:LR1} to understand the sign and approximate strength of the variation of
$\Delta \eta$ with those predictors to first order. 
Therefore, based on the trend of Eq.~\eqref{eq:LR1}, we find that central galaxy alignment effects are
strongest for clusters that are more elongated and higher richness, or clusters that have centrals with larger physical
size, brighter absolute magnitude, redder color, larger ellipticity\footnote{As we will demonstrate
  later in Sec.~\ref{sec: BCG e}, the dependence on central galaxy ellipticity is actually more complicated
  than this simple linear regression result indicates.}, and a higher centering probability.

\subsection{Linear regression: angular segregation of satellite galaxies}
\label{subsection:LR BCG-sat}

In Sec.~\ref{subsection:LR BCG-DM}, we used the positions of satellite galaxies weighted by their
membership probabilities to trace the cluster and underlying halo shape, without any consideration
of individual satellite properties. However, satellite galaxies with different properties are known
to be distributed in different ways within clusters, a phenomenon known as segregation
\citep[e.g.,][]{vdBosch16}. Segregation is often discussed in terms of the radial direction to the
cluster center. Here we investigate angular segregation with respect to the central galaxy major axis, to
understand what satellite properties most strongly predict the satellite tendency to lie along the
central galaxy major axis. 

\begin{figure*}
\begin{center}
\includegraphics[trim=0.15cm 0cm 1.0cm 0cm,width=1\textwidth]{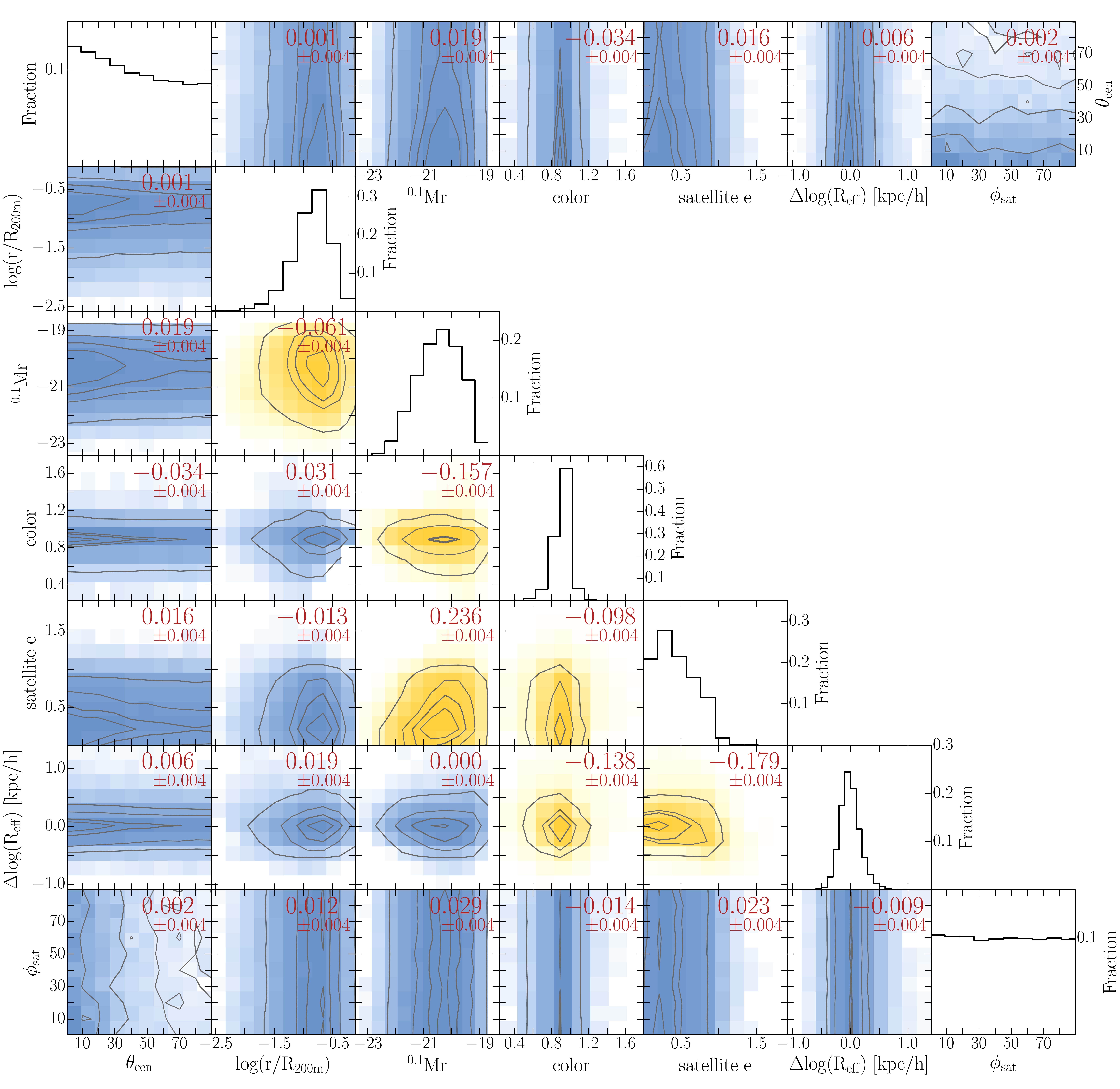}
\caption{Scatterplot matrix of the central galaxy alignment angle with the six satellite related quantities. The correlation coefficient between each
  pair of parameters is noted on the plot. We highlight scatterplots with correlations that are significant at $>10\sigma$ in yellow. 
  The gray contour levels indicate 20\%, 40\%, 70\%, and 95\% number of satellites of our data.}
\label{fig:dr8_satellite_73146}
\end{center}
\end{figure*}

Fig.~\ref{fig:dr8_satellite_73146} shows the scatterplot matrix of $\theta\rm_{cen}$ versus the six
satellite-related quantities for the 73146 DR8 central-satellite pairs.  The top row displays scatterplots between $\theta_{\rm cen}$ and all other satellite quantities. 
Compared with Fig.~\ref{fig:dr8_DM_8233}, the absolute magnitudes of the correlation coefficients of
$\theta_{\rm cen}$ with these satellite quantities are generally smaller than the correlation
between $\Delta \eta$ and central and cluster quantities. 
Although the correlations are weak, the large number of pairs means there is still enough
statistical power to measure these correlations robustly.
In general, the relationships between all pairs of predictors appear to be weak, except for $^{0.1}M_r$ and ellipticity, with a correlation coefficient of 0.236. 
The lower right corner shows the distribution of $\phi_{\rm sat}$, which is very close to flat,
indicating that satellite radial alignment is a far weaker phenomenon compared to central galaxy alignments; we
explore this phenomenon in more detail in future work.

\begin{table}
\caption{Predictors involved in the angular segregation of satellites as analyzed in
  Sec.~\ref{subsection:LR BCG-sat}. The columns are the same as in Table~\ref{tb:predictor_bcg}.
  The bottom panel lists the already-identified central galaxy and cluster quantities. 
  These quantities are included in the linear regression equation when doing variable selection, 
  in order to properly account for their influence on the angular segregation of satellites.
  The top panel shows the four selected satellite quantities that significantly affect angular segregation.}
\begin{tabular}{lrrlrr}
\hline 

Predictor						& $\beta$	& $t$-value	& $p$-value		& mean	& $\sigma$	\\ \hline

satellite color 			 		& -0.66	& -6.7	& $2\times10^{-11}$		& 0.91	& 0.09	\\
satellite $^{0.1}M_r$				& 0.54	& 5.23	& $1\times10^{-7}$		& -20.46	& 0.76	\\
log($r$/R$_{\rm 200m}$)			& 0.21 	& 2.12	& 0.03				& -0.87 	& 0.32	\\
satellite $e$					& 0.20 	& 2.1		& 0.04				& 0.43	& 0.26	\\ \hline

cluster $e$ 					& -2.52	& -25.6	& $2\times10^{-143}$	& 0.21	& 0.11		\\
$\Delta$log(cen.\ R$\rm_{eff}$)		& -0.64	& -6.4	& $1\times10^{-10}$		& 0.02	& 0.15		\\	
cen. $^{0.1}M_r$				& 0.58	& 5.01	& $1\times10^{-6}$		& -22.35	& 0.49		\\
cen. color						& -0.28	& -2.8	& 0.005				& 0.98	& 0.07		\\									
$P_{\rm cen}$					& -0.46	& -4.2	& $2\times10^{-5}$		& 0.87	& 0.17		\\ 
cen. $e$						& -0.42	& -4.2	& $3\times10^{-5}$		& 0.25	& 0.16		\\
log(richness)					& 0.09	& 0.8		& 0.4					& 1.58	& 0.21		\\ 

\hline
\label{tb:predictor_sat}
\end{tabular}
\end{table}

Given that we already identified the important central galaxy- and cluster-related predictors that affect
the central galaxy alignment signal, it is reasonable to include these predictors in our linear
regression analysis in order to  compensate for their influence on the angular segregation of satellites. 
Table~\ref{tb:predictor_sat} shows the results of linear regression,
with the new selected satellite quantities on top and the already-known central galaxy and cluster quantities
on the bottom. Almost all previously-selected quantities have an associated $p-$value below 0.05 when using
$\theta_{\rm cen}$ as the response variable, except for cluster richness $\lambda$. This may be due
to the fact that higher richness clusters tend to be rounder (as revealed in the last row of
Fig.~\ref{fig:dr8_DM_8233}), and thus have their member galaxies less segregated toward any specific
direction. Also, although the regression slope for log(richness) is positive, we cannot infer that
satellites in lower richness clusters tend to be more segregated (i.e. having smaller $\theta_{\rm
  cen}$). The level of angular segregation against richness is not significant enough for us to make
such a conclusion.
The best-fitted linear regression equation with these predictors is

\begin{equation}\label{eq:LR2}
\begin{split}
\theta_{\rm cen} = 41.60 
-0.66\ \frac{{\rm sat.\ color}-0.91}{0.09}	
+0.54\ \frac{{\rm sat.}\ ^{0.1}M_r+20.46}{0.76}			\\
+0.21\ \frac{{{\rm log(}r/{\rm R_{200m}})}+0.87}{0.32}
+0.20\ \frac{{\rm sat.}\ e-0.43}{0.26}						\\
- 2.52\ \frac{{\rm cluster}\ e-0.21}{0.11}					
-0.64\ \frac{\Delta {\rm log(cen}\ R_{\rm eff})-0.02}{0.15} 		\\
+0.58\ \frac{{\rm cen}\ ^{0.1}M_r+22.35}{0.49} 			
-0.28\ \frac{{\rm cen\ color}-0.98}{0.07} 					\\
-0.46\ \frac{P_{\rm cen}-0.87}{0.17}						\\
-0.42\ \frac{{\rm cen}\ e-0.25}{0.16}	 					
+0.09\ \frac{{\rm log(richness)-1.58}}{0.21}
\end{split}
\end{equation}
The tendency of satellites to reside along the central galaxy major axis is strongest for satellites that are
redder, brighter, rounder, and located closer to the central.


\section{The effect of shape measurement method on the central galaxy alignment signal}
\label{sec:shape-align}

In the rightmost panels of Figs.~\ref{fig:Delta_eta_dist} and \ref{fig:BCG_align_ang_dist}, we
compared the distributions of our response variables, $\Delta \eta$ and $\theta_{\rm BCG}$, using
three different shape measurements methods. In both cases, the level of central galaxy alignment measured via
de Vaucouleurs and isophotal shapes agree with each other within the error bar, while the
re-Gaussianization measurement gives us a less strong central galaxy alignment effect. In this section, we
discuss the interpretation of this result in terms of systematic and physical effects in these shape
measurements. Our discussion also relies on results of \citet{Singh15b}, who analyzed the effect of
these three shape measurement methods on the inferred galaxy alignments of luminous red galaxies
(LRGs).

\subsection{Systematic error}

Different shape measurements deal with the effects of the PSF on galaxy images differently. The
re-Gaussianization technique was designed for weak lensing studies requiring the most
complete removal of the PSF effect on galaxy shapes. The de Vaucouleurs shape measurement only
partially corrects for the PSF by using a double-Gaussian fit instead of the full PSF model, while
the isophotal shape measurement does not correct for the effect of the PSF explicitly. 

Another relevant aspect of systematics has to do with what part of the light profile is used for the
measurement.  The re-Gaussianization method has an elliptical Gaussian weight function, emphasizing
the central regions of the profile.  The de Vaucouleurs profile includes both the central region and
the large-scale wings of the light profile, while the isophotal shape measurement {\em only} uses
the 25 mag/arcsec$^{2}$ isophote which is quite far out in the wings.  These choices could make the
latter two methods more sensitive to sky subtraction systematics than the re-Gaussianization method
\citep[for more discussion in the context of the isophotal method, see][]{Hao11}.

One could infer that isophotal shapes would contain severe systematics due to the PSF.  However,
the results of \citet{Singh15b} suggest that the impact of the PSF on the shape measured at very low
surface brightness is quite small. Instead, the de Vaucouleurs shapes exhibited the most significant
systematic errors of the three methods.  
Therefore, we may treat the detected differences in the central galaxy alignment strength between the
re-Gaussianization and isophotal shapes as reflecting a true physical effect that we will discuss
below. However, we should keep in mind that the systematic tests in \citet{Singh15b} were based on a
specific 
sample of galaxies, while our central galaxy sample (which is preferentially located in regions of high galaxy
density) may still suffer from some contamination in the isophotal shapes, as suggested by
\citet{Hao11}.

\subsection{Physical effect}

The higher apparent degree of central galaxy alignment using isophotal shapes compared to that using
re-Gaussianization sahpes may be primarily due to a mechanism called ``isophote twisting''
\citep{diTullio78,diTullio79,Kormendy82,Romanowsky98,Lauer05}. 
The physical origin of this effect is that the outer part of the galaxy light profile may respond
more strongly to tidal fields than the inner part of the galaxy. Thus, by tracing the outermost
isophote of the galaxy, the isophotal shape records the highest level of alignment with the tidal
field.


\section{The origin of central galaxy alignment}
\label{sec: BCG origin}

In Sec.~\ref{subsection:LR BCG-DM} we applied linear regression analysis to the nine central galaxy and
cluster quantities, and picked the predictors that significantly influence the alignment between central galaxy and its host cluster. 
We now address the origin of this alignment phenomenon and compare our results with previous studies. 

\subsection{Dependence on cluster ellipticity}
\label{sec: cluster ellipticity}

Simulations have revealed that clusters are triaxial rather than spherical
\citep{Jing02,Hopkins05,Kasun05,Allgood06,Hayashi07}, so they look elongated when projected on the sky. 
The last panel in the second last row of Fig.~\ref{fig:dr8_DM_8233} shows the distribution of projected redMaPPer
cluster
ellipticities traced by the weighted member galaxy distribution (see Sec.~\ref{subsec:cluster e} for definition of cluster ellipticity), with 
a mean cluster ellipticity 
of $\sim$0.20 and a mean 
projected semi-minor to semi-major axis ratio of $\mean{b/a}\sim0.67$, which
agrees with the $N$-body simulation of \citet{Hopkins05} ($\mean{b/a}\sim0.67$, at redshift zero),
but is rounder than that
directly measured through gravitational lensing ($\mean{b/a}\sim0.48_{-0.09}^{+0.14}$ in \citealt{Evans09}, and $\mean{b/a}\sim0.46\pm 0.04$ in \citealt{Oguri10}). 

As shown in Table~\ref{tb:predictor_bcg}, we find that cluster ellipticity has the most significant
influence on the central galaxy alignment signal, with centrals in more elongated clusters having a stronger
alignment with the orientation of their host clusters (see also the second-to-last panel in the first row of
Fig.~\ref{fig:dr8_DM_8233}, which directly displays the correlation between cluster ellipticity and
central galaxy alignment). 
Since the position angle for round clusters is not very meaningful, particularly given observational
noise, we have examined the correlation trend for clusters with ellipticity $> 0.2$, and found the
trend that more elongated clusters show stronger alignment still holds. 

\begin{figure*}
\begin{center}
\includegraphics[width=\textwidth]{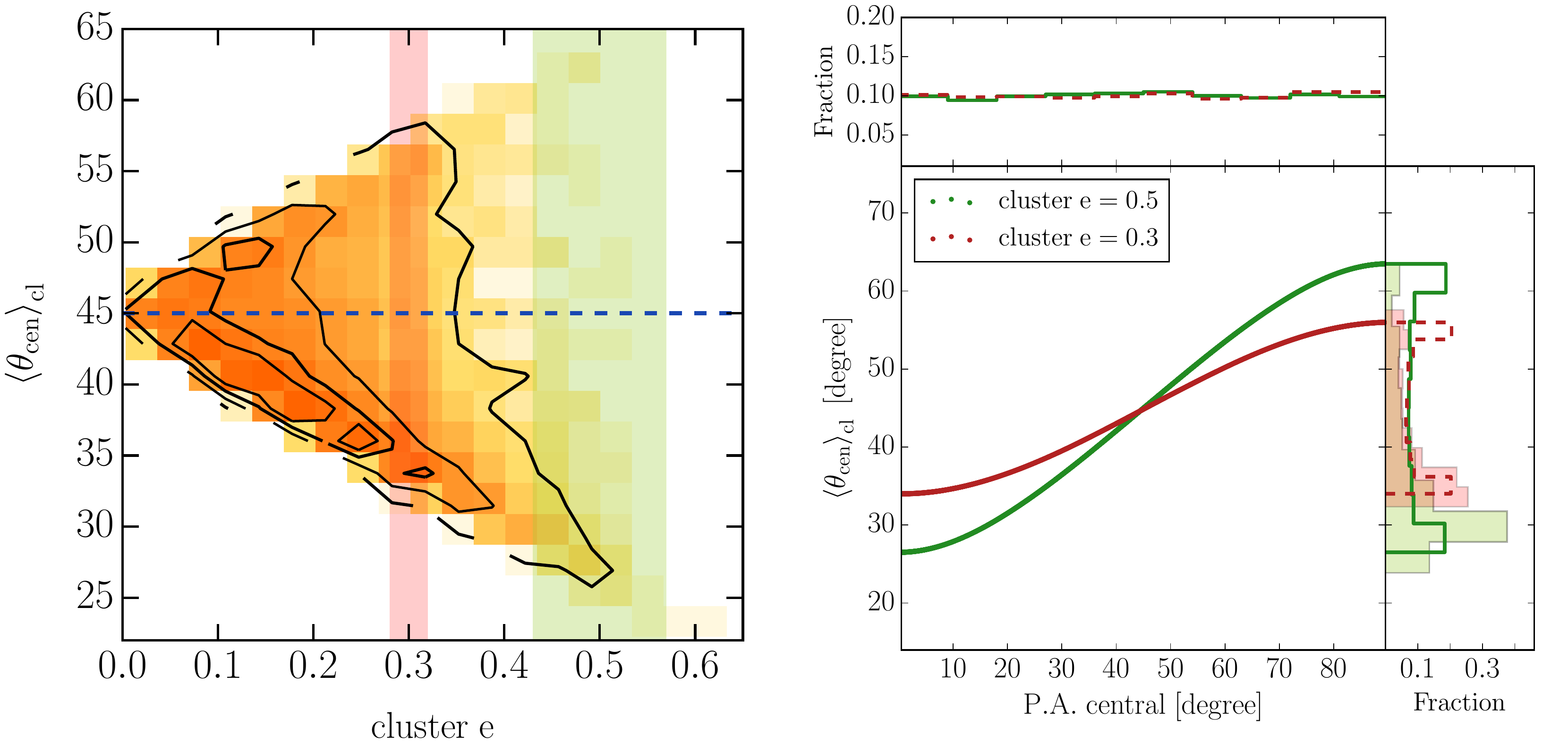}
\caption{Left Panel: Contour plot between cluster ellipticity and averaged central galaxy alignment
  angle for all central-satellite pairs in each cluster, $\mean{\theta_{\rm cen}}_{\rm cl}$. The
  blue dashed line indicates the case $\mean{\theta_{\rm cen}}_{\rm cl} = 45^{\circ}$, when
  satellites are randomly distributed within cluster.   
  The light-green (pink) shaded area marks out clusters with ellipticity in the range of 0.43$\sim$0.57 (0.28$\sim$0.32). 
  For each cluster ellipticity value, there are
  more clusters distributed in the region below the blue dash line than above, showing the tendency
  for central galaxy alignments. 
Right Panel: Non-linear relationship between central galaxy  P.A.\ and the derived $\mean{\theta_{\rm
    cen}}_{\rm cl}$ in our simulated data. The simulated clusters with cluster ellipticity of 0.5
(0.3) are shown in green (red). With completely random distributions of simulated central galaxy
P.A.,  the distributions of derived $\mean{\theta_{\rm cen}}_{\rm cl}$ tend to peak at their minimum or maximum available values.
The light-green and pink shaded histograms are the distributions of $\mean{\theta_{\rm cen}}_{\rm cl}$ in our observational data within certain
cluster ellipticity ranges as highlighted in the left panel. 
}
\label{fig:PavgBCGang_cluster_e}
\end{center}
\end{figure*}

The influence of cluster ellipticity on central galaxy alignment can be further visualized in the left panel of 
Fig.~\ref{fig:PavgBCGang_cluster_e}, where we plot the distribution of the $p_{\rm mem}$-weighted
averaged central galaxy alignment angle for all $p_{\rm mem}>0.2$ central-satellite pairs in each cluster,
$\mean{\theta_{\rm cen}}_{\rm cl} = \frac{\sum_{i} p_{{\rm mem},i} \theta_{\rm cen}}{\sum_{i}
  p_{{\rm mem},i}}$, against the cluster ellipticity. The sharp boundary on each side is due to the
way we define cluster ellipticity. Since we calculate cluster ellipticity via the satellite galaxy
distribution, round clusters  (with satellites distributed in an almost circularly symmetric way) thus
have $\mean{\theta_{\rm cen}}_{\rm cl} \sim 45^{\circ}$. More elongated clusters have more potential
for going to lower or higher $\mean{\theta_{\rm cen}}_{\rm cl}$ values. 
At fixed cluster ellipticity, the distribution of $\mean{\theta_{\rm cen}}_{\rm cl}$ tends to cluster 
towards the edges of the minimum and maximum available values. 
As a demonstration, the right panel of Fig~\ref{fig:PavgBCGang_cluster_e} shows the results of
simulating two fake clusters with fake member galaxies distributed with elliptical symmetry such
that the two clusters would have measured cluster ellipticity of 0.5 (green) and 0.3 (red).
We then randomized the P.A. of the simulated central galaxies, and calculated the corresponding $\mean{\theta_{\rm cen}}_{\rm cl}$ value. 
From the scatter plot and histograms of P.A. central vs.\ $\mean{\theta_{\rm cen}}_{\rm cl}$,
it is clear that the relationship between central galaxy P.A. and $\mean{\theta_{\rm cen}}_{\rm cl}$
is non-linear, and that this non-linearity is responsible for the shape of the left panel of Fig.~\ref{fig:PavgBCGang_cluster_e}. 
However, with more clusters distributed on the $\mean{\theta_{\rm cen}}_{\rm cl} < 45^{\circ}$ side across the full cluster ellipticity range shown in the left panel of Fig~\ref{fig:PavgBCGang_cluster_e}, centrals do prefer to align with their overall satellite distributions.

There are two mechanisms that may be responsible for the strong dependence of central galaxy alignment on
cluster shape: 1) the imprint of infall of matter and galaxies into the cluster preferentially along
filaments, and 2) the large-scale tidal gravitational field (either primordial, at the time of central galaxy
formation, or tidal torquing over time). 
First, centrals and their parent clusters are both formed via accreting galaxies along filaments, which
imprint preferred directions. As a result of these inflows, we expect central galaxies to be aligned with their
clusters, especially for relatively young and small clusters with only one dominant filament,
leaving an elongated distribution of galaxies \citep{Knebe04, Libeskind05, Libeskind15}. 
More massive clusters may have experienced several merger events along filaments in various
directions during their assembly history. This more complicated history makes the distribution of
galaxies in these clusters more disturbed and randomized, resulting in a rounder shape. Indeed, as
shown in the second-to-last row of Fig.~\ref{fig:dr8_DM_8233}, there is a weak anti-correlation between
cluster ellipticity and richness in our data, with richer clusters having a smaller ellipticity. 
The subsequent violent merger activities may wash out the memory of the primordial
filamentary structure, causing a reduction in the alignment signal \citep{Ragone07}. 

However, over the process of virialization, the distribution of galaxies in clusters would again
gradually be stretched out along the direction with the surrounding large-scale tidal field,
reaching new equilibrium states with a triaxial morphology. At the same time, central galaxies would also
gradually be tidally torqued along the new established direction of tidal field. A more anisotropic
distribution of satellites could indicate a more intense tidal fields to torque the centrals. 
It is unclear how important this instantaneous torquing is; \citet{Camelio15} demonstrated that at
galaxy scales, it is too weak to account for the observed intrinsic alignments, but it is unclear
whether it is definitely subdominant for cluster mass scales.

We emphasize that the above two scenarios (anisotropic infall and tidal torquing) are not mutually
exclusive and do not necessarily have some sequence in time.  They could both operate at various
stages of the cluster and central galaxy evolutionary process.   
Also, according to the linear alignment model \citep[e.g.,][]{Catelan01, Hirata04a}, the intrinsic
alignment is already set by tidal fields at the time of galaxy formation, and it is not clear how
relevant these additional processes that operate later may be.

\subsection{Dependence on central galaxy effective radius}

As we have shown, the central galaxy effective radius at fixed intrinsic luminosity is also a very significant predictor of the central galaxy alignment
effect, with larger-sized centrals at a given luminosity exhibiting a stronger degree of central galaxy alignment than smaller-sized centrals. 

Observations and semi-analytic models have revealed that most massive galaxies grow inside-out, with
their extended stellar halos dominated by accreted stars. The supply of accretion stars may
originate from the stellar streams \citep{Belokurov06} or the diffuse intracluster light (ICL) which
is composed of tidally-stripped stars that are gravitationally bound to the cluster potential \citep{Oemler76,Lin04}. 
These massive accretion-dominated galaxies thus tend to have more extended light profiles compared
to galaxies with a stellar component that primarily underwent ``in situ'' star formation \citep{vanDokkum10,Cooper13}. 

There are two scenarios that can explain the dependence of central galaxy alignment on central galaxy size. First,  centrals
with more extended morphology may respond more strongly to tidal forces (either the primordial or
instantaneous tidal field). Defined as the difference between the gravitational forces at two
different positions on an object, the strength of the tidal force would be stronger for objects that
have a larger spatial extent. 
The alternative explanation stems from the closely linked formation and evolution histories of centrals
with their host clusters and the surrounding large-scale structures \citep{Conroy07}. As reported in
\citet{Zhao15}, centrals with extended cD envelopes tend to have larger $R_{\rm e}$, and are believed to
be dominated by baryons from accretion. If there is an abundant supply of accretion stars in some
direction aligning with the overall distribution of member galaxies, the central galaxy shape would naturally
extend towards the preferred direction of accretion, and thus align with the angle of the member
galaxy distribution.  We are unable to distinguish between these two scenarios.

\subsection{Dependences on central galaxy luminosity, dominance and centering probability}

According to Fig.~\ref{fig:dr8_DM_8233}, central galaxy luminosity, dominance and centering probability are
mutually highly correlated with each other, and thus are likely caused by similar physical
origins. Here we discuss the dependences of central galaxy alignment on these three predictors.

As revealed in Table~\ref{tb:predictor_bcg}, we found that $\Delta \eta$ depends significantly on
central galaxy $^{0.1}M_r$ and  $P_{\rm cen}$, with centrals that are more luminous and have a higher centering probability 
tending to be more aligned with the cluster position angle. Central galaxy dominance, however, was not selected
as a featured predictor. This does not mean that
central galaxy dominance is not important, but rather that its effect on central galaxy alignment may have been soaked up
by the effects of central $^{0.1}M_r$ and $P_{\rm cen}$, so that knowing the central galaxy dominance provides no
further help when predicting $\Delta \eta$ if the other two predictors are also known. 

Our result is consistent with that of \citet{Hao11}, who also detected a strong dependence of BCG
alignment on BCG luminosity based on a sample of richness $\geq15$ clusters taken from GMBCG, a cluster catalog constructed based on the red-sequence method \citep{Hao10}. 
Also, due to the tight correlation between central galaxy $^{0.1}M_r$ and dominance, with more luminous
centrals showing higher degree of central galaxy dominance, our result implies that clusters with more dominant
centrals should have stronger central galaxy alignment. This agrees with the result of
\citet{Niederste-Ostholt10}. They found that BCG-dominant clusters exhibit 
stronger BCG alignments than less BCG-dominant clusters do, with a difference significant at the 4.4$\sigma$ level, based on both the maxBCG cluster catalog \citep{Koester07a} and a
matched filter cluster catalog of \citet{Dong08}.

The dependences of the central galaxy alignment signal on central $^{0.1}M_r$, dominance and $P_{\rm cen}$ have their common origin in the following aspects. 
1) It may originate from the purity of measurement. Luminous and dominant centrals have a higher
probability of sitting closer to the true center of their dark matter potential wells \citep{Wen13}.
This kind of system suffers less contamination from wrong detections, and could therefore end up
showing a higher central galaxy alignment signal. 
2) Clusters with luminous and dominant centrals are typically more relaxed.
More relaxed systems have experienced the uninterrupted (by mergers) influence of surrounding large-scale tidal fields for a
longer period of time, and thus it may be more likely for their centrals to align. 

\begin{figure*}
\begin{center}
\includegraphics[trim=0.15cm 0cm 1.0cm 0cm,width=1\textwidth]{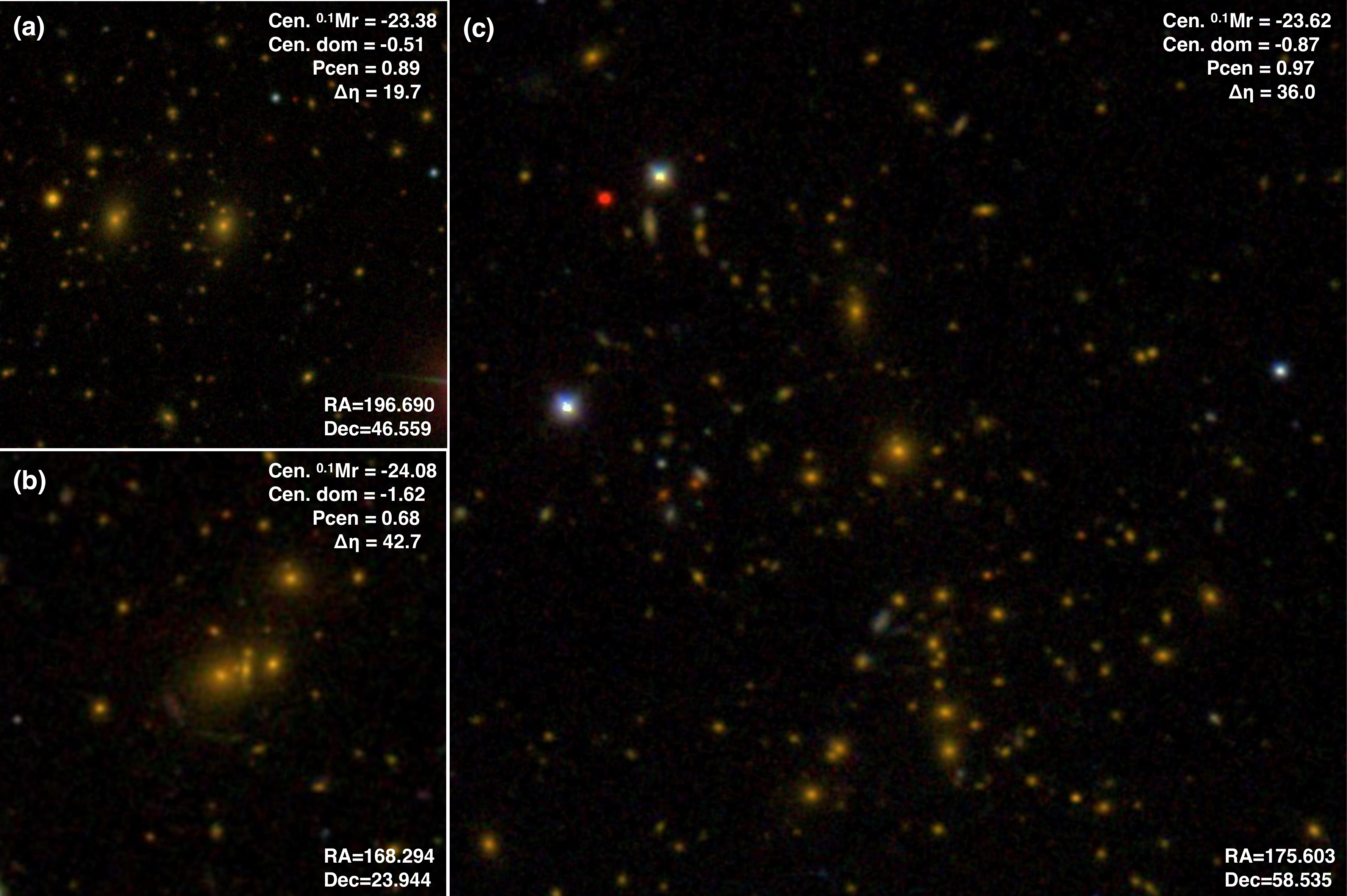}
\caption{Examples of clusters with central galaxies that are luminous but not dominant. The widths of panels (a), (b), (c) are set to be $\frac{1}{4}$, $\frac{1}{5}$, and $\frac{1}{2}R_{\rm 200m}$ (respectively) of their host clusters.}
\label{fig:coma_like}
\end{center}
\end{figure*}

Given that central $^{0.1}M_r$ and dominance are highly correlated at $\sim$0.6, it is natural to
ask what causes us to select central $^{0.1}M_r$ rather than dominance as a featured predictor? To
address this question, in Fig.~\ref{fig:coma_like}, we show some example clusters with luminous but less dominant centrals. 
As shown, these clusters typically have several bright galaxies, and may still be undergoing
significant merging and disruptive interactions. 
Figs.~\ref{fig:coma_like}a and c show examples of clusters with their dominantly bright members
still some distance away from the centrals. These systems may be not relaxed, but if the centrals' high
luminosities and the distributions of their members stem from the same primary avenue of accretion, 
high alignment signals can still shown even if the centrals are not dominant. This explains why the importance of central $^{0.1}M_r$ stands out from central galaxy dominance.

Fig.~\ref{fig:coma_like}b shows the case where the bright members already sank into the potential
well of the cluster and are closely interacting with the central galaxy. In this case, the orientation of the
central galaxy may be affected temporarily by these closely interacting galaxies, rather than reflecting the
tidal field originating from the large-scale environment. The upper right corner of each panel in
Fig.~\ref{fig:coma_like} shows some physical properties of the central galaxy. In the case of
Fig.~\ref{fig:coma_like}b, with several bright galaxies crowded in the central region of the
cluster, the central $P_{\rm cen}$ tends to be low. This demonstrates that $P_{\rm cen}$ can still be
selected as a featured predictor even after selecting central $^{0.1}M_r$, because it indicates whether
there are other bright galaxies near the central that may reduce the central galaxy alignment with the large-scale
tidal field through dynamical processes.  
While examining images of individual clusters does not give the full picture, it is a way of
supplementing the statistical measure of central galaxy alignment from the linear regression analysis.
 
\subsection{Dependence on central galaxy color}
\label{sec:BCG_color}

We observed that redMaPPer centrals with redder color show stronger central galaxy alignments. The
enhancement of the central galaxy alignment signal among red hosts has also been observed in systems across a wide
range in halo masses. Based on a sample of isolated host galaxies with typically $1$--$2$
csatellites, \citet{Azzaro07} and \citet{Agustsson10} found an excess of
satellites along the major axis of their centrals only in red-colored hosts, while satellite
distributions are consistent with isotropic around blue hosts. Based on group catalogs spanning from
isolated host to cluster scale halos, \citet{Yang06}, \citet{Wang08}, and \citet{Siverd09} all 
found that the alignment signal is only detected in groups with red centrals, and is strongest when
considering red centrals and red satellites.  

Unlike those previous works, our sample is selected based on
the red-sequence method, so the centrals all belong to the red galaxy population. Within the red
population, we nonetheless found that the central galaxy alignment depends on the $^{0.1}Mg-^{0.1}Mr$ color of
central galaxy in cluster scale. Galaxy color indicates the age of the stellar populations. Recent star
formation activities induced by the supply of gas from surrounding materials or merger events would
cause the central galaxy color to become less red. Our result thus suggests that central galaxy alignment signal preferentially exists in centrals with relatively old stellar population. For clusters with bluer central galaxies, the alignment of centrals may be disturbed by the recent merger events that also triggered star formation and
contributed to the bluer color. 

\subsection{Dependence on central galaxy ellipticity}
\label{sec: BCG e}

Our linear regression shows that central galaxy ellipticity (as defined in Eq.~\ref{eq:e}) is negatively correlated with $\Delta \eta$, which
means that centrals with larger ellipticity exhibit stronger central galaxy alignment. However, the complication in
detecting this trend is that it is more difficult to accurately determine the position angles for
round centrals. Statistical scatter in measuring the position angles of more round centrals could in principle drive the effect we
have observed, rather than it being a true physical effect.  Many studies have required the central galaxy
ellipticity to exceed some value in order to avoid this effect, at the expense of introducing some 
systematic selection effect.  

\begin{figure*}
\begin{center}
\includegraphics[trim=0.15cm 0cm 1.0cm 0cm,width=1\textwidth]{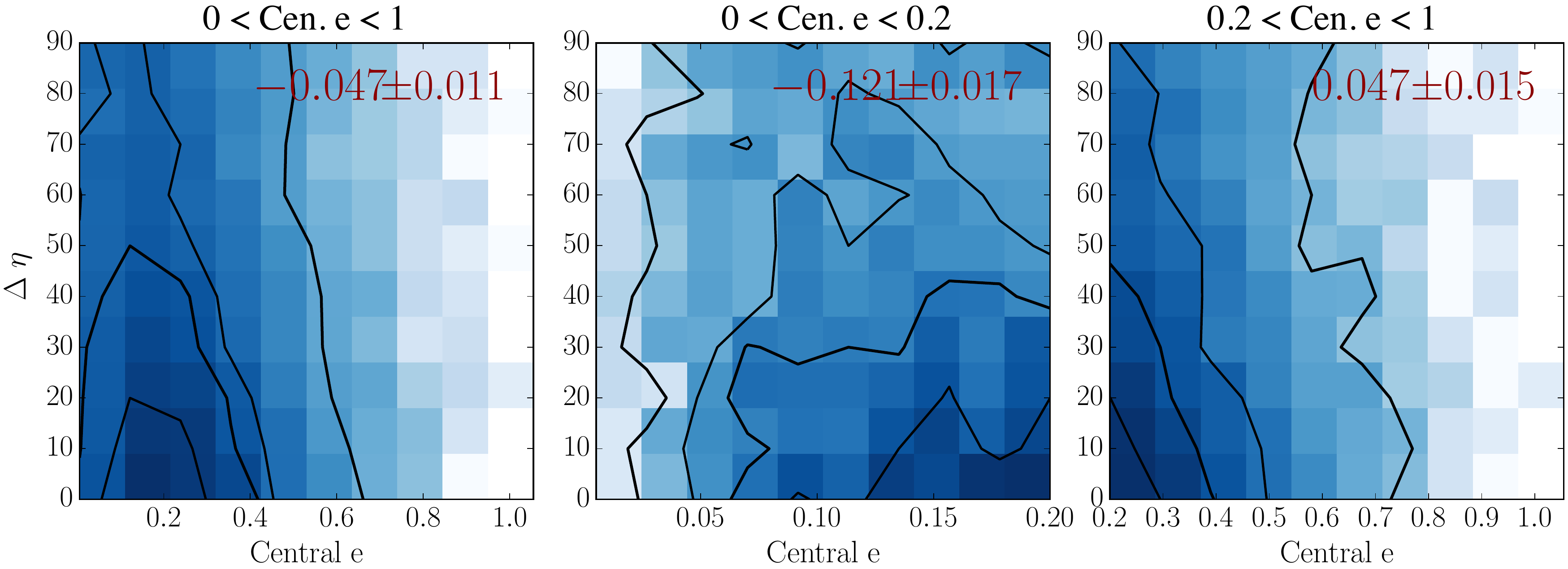}
\caption{Contours of scatterplots of central galaxy ellipticity v.s. the position angle difference between central galaxy and cluster ($\Delta \eta$) in different central galaxy ellipticity bins. The left panel shows all of our cluster sample, the middle panel shows only clusters with central galaxy ellipticity below 0.2, while the right panel plots clusters with their central galaxy ellipticity above 0.2. The correlation coefficient between $\Delta \eta$ and central galaxy ellipticity is shown at the upper right corner in each panel. }
\label{fig:BCG_e}
\end{center}
\end{figure*}

To address this issue, in Fig.~\ref{fig:BCG_e} we show what happens to the correlation between central galaxy
ellipticity and central galaxy alignment angle when we divide the original full sample (left panel) into two
ellipticity bins at a value of 0.2, with 3554 centrals in the < 0.2 bin, and 4679 centrals in the other.              
This division reveals that the detected negative correlation of -0.047 in the full cluster sample is
dominated by centrals with ellipticity below 0.2 (middle panel), in which a correlation coefficient of
-0.12 is measured. These centrals are particularly sensitive to measurement error in the position angle,
so the observed negative correlation may arise at least in part from measurement error.  
It may be also possible that this negative correlation originates from real physical mechanisms. The
morphology of centrals reflect their formation history. More elliptical centrals may have experienced more
anisotropic accretion that contributes to a stronger alignment effect.  Distinguishing between
measurement error and this real physical effect is difficult.

If focusing on systems with ellipticities above 0.2 (right panel), the central galaxy alignment angle becomes
positively correlated at $0.044\pm 0.015$, meaning that more elongated centrals have smaller alignment
signals. We also find that the observed positive correlation is largely driven by the 8\% highest ellipticity centrals, with ellipticity $\geq$ 0.5.
Our result agrees with that of \citet{Yang06} (see their Fig.~2), who found the same
tendency using groups with central galaxy ellipticity\footnote{The definition of galaxy ellipticity adopted in \citet{Yang06} is 1-b/a, based on SDSS isophotal measurement.} 
$\geq$ 0.2.  

\begin{figure*}
\begin{center}
\includegraphics[trim=0.15cm 0cm 1.0cm 0cm,width=0.75\textwidth]{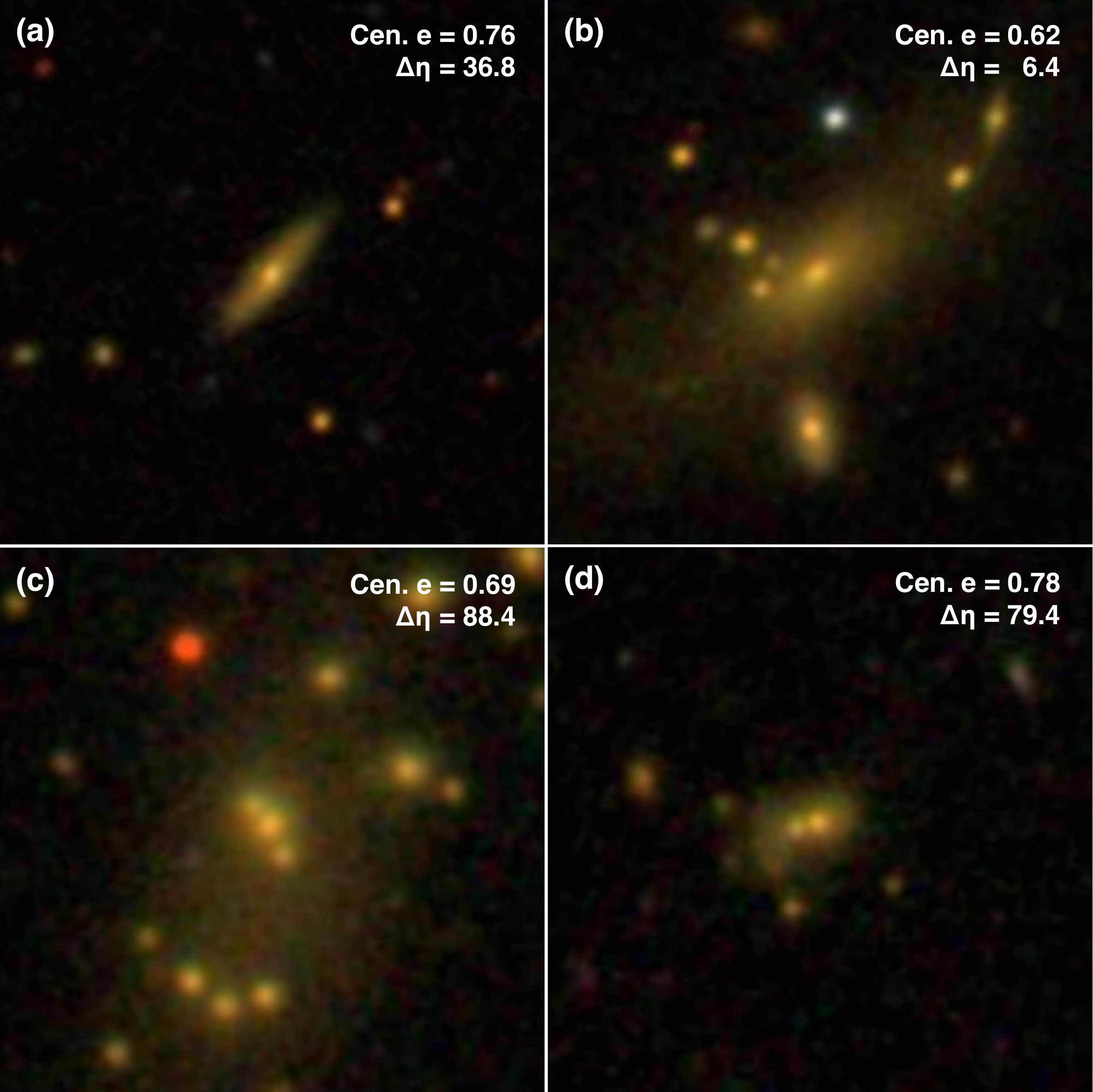}
\caption{Examples of centrals with measured ellipticity $\geq$ 0.5. The central galaxy ellipticity and position
  angle difference between the central galaxy and cluster member galaxy distribution ($\Delta \eta$) is shown
  in the upper right corner of each panel. All panels are 150 kpc on each side. (a) Disky structure central with blueish color. 
  (b) Central galaxy with elongated ICL. (c)$\&$(d) Centrals with double or more bright cores within common extended envelopes.}
\label{fig:BCG_e_SDSS}
\end{center}
\end{figure*}

What causes high-ellipticity centrals to be less aligned? To partially address this question, we
visually inspected the images of centrals with very high ellipticities ($\geq$ 0.6) and presented some
examples in Fig.~\ref{fig:BCG_e_SDSS}. 
Surprisingly, besides the expected cases of high-ellipticity centrals that are more blue and exhibit
disky structures (Fig.~\ref{fig:BCG_e_SDSS}a) or those with anisotropic ICL
(Fig.~\ref{fig:BCG_e_SDSS}b), we found that in many instances, high-ellipticity centrals are systems
with $\ge 2$ bright cores in a single extended envelope (Fig.~\ref{fig:BCG_e_SDSS}c, \ref{fig:BCG_e_SDSS}d). These
multiple-core centrals are currently undergoing mergers. During the violent coalescence processes, the
position angles of centrals change rapidly and no longer reflect the large-scale matter distribution,
resulting in a wide spread in $\Delta \eta$. 

We conclude that we should ignore central galaxy ellipticity as an predictor, although it is significantly identified through our variable selection process. The observed negative correlation is mostly driven by rounder centrals whose P.A. determination is more likely affected by systematics. For more elongated centrals, positive correlation with $\Delta \eta$ is found, and this correlation is possibly driven by centrals at higher ellipticity end. So far we cannot draw a clear conclusion about the impact of central ellipticity on
central galaxy alignments.
Larger sample size and improved shape measurement method in the future would help us to analyze the non-linear relation between $\Delta \eta$ and central galaxy ellipticity.

\subsection{Dependence on richness}

Richness was selected as a statistically significant predictor when using $\Delta \eta$ as the
response variable, but not for the response variable $\theta\rm_{cen}$. This result suggests that
the impact of richness on the central galaxy alignment signal is marginal. We refer the reader back to the ending of Sec.~\ref{subsection:LR BCG-sat} for related discussion.
At similar cluster mass scales, \citet{Niederste-Ostholt10} also found a slight indication that richer clusters show stronger alignment signals, at 2.3$\sigma$ significance,
while \citet{Hao11} detected no dependence of BCG alignment on richness.

Observationally, richness is a good estimator for the underlying cluster dark matter halo mass \citep{Rykoff12}. The weak dependence on richness may be due to the limited range of halo masses covered by the redMaPPer cluster sample. 
In what follows, we compare papers in which the mean central galaxy alignment angles, $\mean{\theta\rm_{cen}}$, are provided, and summarize the comparison results in Table~\ref{tb:compare}. Since almost all of the previous works used the isophotal shape measurements, we also turn to our isophotal measurements to fairly compare the $\mean{\theta\rm_{cen}}$ values.
At the mass scale corresponding to galaxy groups, many studies have observed that there is a
 stronger alignment tendency in richer groups \citep{Yang06, Wang08, Siverd09}. As shown in
 Table~\ref{tb:compare}, the $\mean{\theta\rm_{cen}}$ value in the highest mass bin of
 \citet{Yang06} is consistent with our isophotal $\mean{\theta\rm_{cen}}$. 
Going down to even smaller systems, \citet{Brainerd05} and \citet{Agustsson10} have measured the
$\mean{\theta\rm_{cen}}$ using a sample of isolated host centrals. The values of
$\mean{\theta\rm_{cen}}$ are generally larger than that measured in cluster scales. 
Therefore, we suggest that there truly is some effect of host halo mass on alignments, despite our
marginal findings using richness
as a mass tracer on cluster mass scales.

Another possible reason that richness may be a less significant predictor is due to the cluster assembly process.
Perhaps originally more massive and richer clusters had a stronger primordial alignment with the tidal field, 
but the subsequent mergers and other major events washed them out, making central galaxy alignments
depend only weakly on richness.

\begin{landscape}

\begin{table}
\caption{
Summary of central galaxy alignment measurements. Here we provide a detailed comparison of the observed average
central galaxy alignment angle, $\mean{\theta\rm_{cen}}$, from previous work using a variety of datasets in
order to test for the potential evolution of $\mean{\theta\rm_{cen}}$ with halo mass. Relevant properties of the sample used in these studies are also listed. 
}
\begin{tabular}{lcccccl}
\hline 

paper                                    &   $\mean{\theta\rm_{cen}}$ & redshift & \thead{shape \\ measurement \\ method }    & \thead{SDSS \\ pipeline }& N\lowercase{pair}  & cluster catalog properties    \\ \hline  \hline
This work DR8\ re-Gaussianization      &  41.42$\pm$0.08          &               & re-Gaussianization    &  DR8               & 94817   &  $\bullet$ redMaPPer      						    \\   
\ \ \ \ \ \ \ \ \ \ \ \ \ \ DR7\ de\ Vaucouleurs  &  40.74$\pm$0.09         & 0.1-0.35 &  de Vaucouleurs       &   DR7               & 86350   & 	$\bullet$ halo mass $\gtrsim 10^{14}$ $h^{-1}$M$_\odot$ 	     \\
\ \ \ \ \ \ \ \ \ \ \ \ \ \ DR7\ isophotal    &  40.60$\pm$0.09         &               &  isophotal         &   DR7              & 86350    & 	$\bullet$ richness $\gtrsim 20$ 		  			     \\ \hline

\citet{Yang06}   						  	 &				&		&		&		    	    &		   &  $\bullet$ \citet{Weinmann06} group catalog      \\ 
\ \ \ \  log[$M_{\rm halo}$/($h^{-1}$M$_\odot$)]:12-13		  & 43.1$\pm$0.4    & 		&		& 			    & 		   &   \ \ \ based on the group finder of \citet{Yang05a}  \\
\ \ \ \  log[$M_{\rm halo}$/($h^{-1}$M$_\odot$)]:13-14            & 42.6$\pm$0.3    &  0.01-0.2 &  isophotal          & DR2		    & 24728  &   $\bullet$ 9220 binary, 3073 triplet, 3270 member > 3 groups\\
\ \ \ \  log[$M_{\rm halo}$/($h^{-1}$M$_\odot$)]:14-15		  & 40.7$\pm$0.5    & 		&		& 			    & 		   &   $\bullet$ restricted to groups with BCG ellipticity > 0.2  \\  \hline

\citet{Wang08} \\
\ \ \ \  log[$M_{\rm halo}$/($h^{-1}$M$_\odot$)]:12-13	 & 43.38$\pm$0.15    	& 		  &		& 			    & 		   &        \\ 
\ \ \ \  log[$M_{\rm halo}$/($h^{-1}$M$_\odot$)]:13-14     & 42.41$\pm$0.13    	&  0.01-0.2 &  isophotal      & DR4		    & 62212  &   $\bullet$ \citet{Yang07} group catalog \\
\ \ \ \  log[$M_{\rm halo}$/($h^{-1}$M$_\odot$)]:14-15	 & 41.25$\pm$0.41    	& 		  &		& 			    & 		   &    \\  \hline

\citet{Brainerd05} Sample1 & 42.1$\pm$0.5  	& $z\rm_{median}$: 0.05 & isophotal & DR3 	    & 3292 	    &   \makecell[l]{ $\bullet$ isolated host galaxy  \\ 
																			         $\bullet$ dominated by systems containing 1$\sim$2 satellites} \\ \hline
\citet{Agustsson10} 		  & 42.9$\pm$0.5 	& 0.01-0.15			& isophotal & DR7	& 7399 & \makecell[l]{ $\bullet$ isolated host galaxy  \\ 
																			         \ \ \ same selection criteria as Sample1 of \\
																			         \ \ \ \citet{Brainerd05}} \\ \hline
\label{tb:compare}
\end{tabular}
\end{table}

\end{landscape}

\subsection{Dependence on redshift}

We did not find any significant redshift dependence of central galaxy alignment within the limited redshift range of 0.1--0.35. 
In agreement with our observation, \citet{Kang07} studied the alignment strength from redshift 2 to
0 and found no redshift evolution based on $N$-body simulations with a semi-analytical model for
galaxy formation. Based on hydrodynamic simulations, \citet{Tenneti15b} showed a weak redshift dependence
on the intrinsic alignment amplitude at galaxy mass scales, with the alignment signal decreasing at lower redshift.
However, based on samples at cluster scale, both \citet{Niederste-Ostholt10} and \citet{Hao11} have
found that the BCG alignment signal is stronger as redshift decreases within the redshift ranges of
$0.08 < z < 0.44 $ and $z<0.4$, respectively. The discrepancies between their results and ours may
arise from the following: 1) The two previous studies have considered slightly wider redshift ranges
than us such that the redshift-dependent trends become detectable. 2) The observed redshift
evolution may be just a reflection of possible combined evolutions with other physical predictors,
since those two studies did not consider as many parameters as we do. 3) For studies that based on
isophotal shape, there may be more contamination from systematic errors at lower redshift,
since for an apparently brighter BCG (at fixed luminosity), its 25 mag/arcsec$^{2}$ isophote traces
a larger radius where the light of BCG is more easily confused with that from other neighboring
satellites. For our redMaPPer sample, when using isophotal shape measurements, we find that the
correlation coefficient between $\Delta \eta$ and $z$ is $\sim$1.5 times higher than that based on
re-Gaussianization shape. As discussed in Sec.~\ref{sec:shape-align}, this could be partly due to
a systematic and partly driven by a real physical effect.

Studying the redshift evolution of the overall central galaxy alignment signal is important for understanding
the physical mechanism that is responsible for it. If the central galaxy alignment largely stems from the primordial tidal field at the
time of cluster formation \citep{Catelan01, Hirata04a}, later merging or virialization processes may
weaken the primordial signal \citep{Hopkins05}. 
However, if the central galaxy alignment is dominated by signals established from underlying tidal fields
acting during the entire lifetime of clusters, or as suggested by \citet{Niederste-Ostholt10}, the
primordial alignment signals could be enhanced by the secondary infall episodes, we may expect
stronger alignment toward lower redshifts. Currently we lack data to make a convincing conclusion
about redshift evolution of central galaxy alignment; further simulations or deeper observational data pushing
to higher redshift are needed to further investigate this problem.

\subsection{Dependence on cluster concentration $\Delta_{\rm R}$}

\citet{Miyatake16} observed that separating redMaPPer clusters with similar richness and
redshift distributions into large-$\overline{R}_{\rm mem}$ and small-$\overline{R}_{\rm mem}$
populations (see Eq.~\ref{eq:Rmem} for definition of $\overline{R}_{\rm mem}$) yields two cluster
subsamples with similar halo masses, but different large-scale biases. Based on the $N$-body
simulation in the work of \citet{More16}, $\overline{R}_{\rm mem}$ is found to be a good indicator
for cluster mass accretion rate. \citet{Miyatake16} thus interpreted the detected difference in
large-scale bias as evidence for halo assembly bias, wherein the clustering of halos depends not only on their mass, but also on other properties related to their assembly histories, such as halo formation time, mass accretion rate, concentration, and spin (see, e.g., \citealt{Gao05, Wechsler06, Gao07, Dalal08, Lin16}). 
Regardless of whether this result indicates assembly bias or some other physical effect can explain
the differences in large-scale bias, $\overline{R}_{\rm mem}$ does correlate with the concentration of
the cluster member galaxy distribution, and it is nonetheless interesting to test whether $\overline{R}_{\rm mem}$ influences central galaxy
alignments. Here we use the parameter $\Delta_{\rm R}$, which removes the richness and
redshift dependence of the observed concentration of the member galaxy distribution (Eq.~\eqref{eq:DeltaR}).

We found that $\Delta_{\rm R}$ has no effect on the central galaxy alignment. In fact, the correlation coefficient between $\Delta \eta$ and $\Delta_{\rm R}$ is the smallest (0.016) among our predictors, as shown in the upper right corner of Fig.~\ref{fig:dr8_DM_8233}. 
Moreover, the last row of Fig.~\ref{fig:dr8_DM_8233} shows that $\Delta_{\rm R}$ does not have any
$> 10\sigma$ correlations with other parameters, and is therefore relatively independent from the
rest of the parameter space considered in this work.


\section{The origin of angular segregation of satellites}
\label{sec: satellite origin}

We find that the angular segregation of satellites with respect to their central galaxy major axis direction
depends strongly on satellite color and $^{0.1}M_r$, and weakly but still significantly 
on log($r/R_{\rm 200m}$) and satellite ellipticity, as shown in Table~\ref{tb:predictor_sat} in
Sec.~\ref{subsection:LR BCG-sat}. In the following we discuss the possible origins of these dependencies, and compare
our results with previous work.  
We remind the reader that instead of considering all satellite galaxies, our analysis is only based
on red-sequence satellites with membership probability above 0.8 according to the redMaPPer algorithm. 


\subsection{Dependence on satellite color}
\label{subsection: sat color}

The color of the red-sequence satellites is the strongest predictor of their angular segregation,
with redder satellites tending to preferentially lie along the major axis direction of centrals. 
This result agrees with previous work that considered satellites in a wider color range and revealed
that the distribution of redder satellites shows more anisotropy than that of bluer ones \citep{Yang06, Azzaro07, Faltenbacher07, Wang08, Agustsson10}. 

Part of the dependence on satellite color may originate from galaxy properties in filaments
connected to clusters. Clusters assembled mainly by accreting satellites from surrounding
filaments \citep[e.g.,][]{Onuora00, Lee07}. As a result, galaxy properties in filaments may leave some imprint on substructures
within clusters. Using a filament catalog \citep{Chen15c} constructed from SDSS, \citet{Chen15d}
found that red galaxies are on average closer to filaments than blue galaxies. Hence, the observed angular
segregation of redder satellites may be due to their being preferentially accreted along filaments,
which likely have more tendency to align with the major axes of centrals (see also \citealt{Kang07}). 
 
Another possible explanation for the angular segregation by color is related to environmental
quenching. Galaxies in denser environments are redder than galaxies of similar mass in less dense
environments \citep{Peng10, Peng12}. Thus, satellites falling along denser filamentary channels
would tend to be redder than those falling into the cluster from the field \citep{Martinez16}. Those
falling into the cluster along filaments already pre-quenched. Also, satellites orbiting closer to
the major axis direction of centrals should experience higher environmental quenching efficiency due to the higher matter density there.

\subsection{Dependence on satellite luminosity}
\label{subsection: sat lum}

More luminous satellites are more likely to lie along the major axis directions of centrals. For
isolated host-satellite systems, \citet{Agustsson10} have also found a consistent trend. 

The satellite luminosity dependence may also have its origin from galaxy properties in filaments. By analyzing galaxies in filaments, \citet{Chen15d} found that more massive galaxies tend to be closer to filaments than lower mass galaxies. \citet{Li13} observed that there is a significant alignment between the orientations of brightest satellite galaxies with the major axes of their groups, suggesting that brightest satellite galaxies entered their host groups more recently than other satellites.
Using $N$-body simulations, \citet{vdBosch16} have also shown that subhalos with a larger mass at
the time of accretion (a quantity used to link with galaxy stellar mass through abundance matching)
tend to be accreted at a later time (see their Fig.~5) with smaller orbital energy (i.e., on more bound orbits, see their Fig.~9). 
Combining these previous findings, the physical picture is that more luminous satellites are more likely
in-falling from filaments connected to clusters. Since they are accreted by the cluster at a later time, they have not yet
orbited enough to lose the imprint of their original large-scale structure. Furthermore, with
smaller orbital energy at infall, their dynamics would be more easily influenced by the overall mass
distribution in the cluster, and thus as they settle into orbit in the cluster potential well they
are more likely to remain along the major axis direction of the central galaxy. 

\subsection{Dependence on satellite-central distance}

We found that the satellite-central distance is a statistically significant predictor of the angular
segregation of satellites, with those closer to centrals being more likely to be located along the major
axis directions of central galaxies. This may seem puzzling given that the second panel in the first row of
Fig.~\ref{fig:dr8_satellite_73146} shows that the correlation coefficient between log($r$/R$_{\rm
  200m}$) and $\theta_{\rm cen}$ is consistent with zero within the error bar. Apparently,
log($r$/R$_{\rm 200m}$) is selected as a feature predictor due to some interplay with another predictor. 
To identify which other predictor is responsible, we removed one predictor at a time in
Eq.~\eqref{eq:LR2} to find which one, when removed, caused log($r$/R$_{\rm 200m}$) to no longer
be selected as a feature predictor. 

\begin{figure}
\begin{center}
\includegraphics[width=0.35\textwidth]{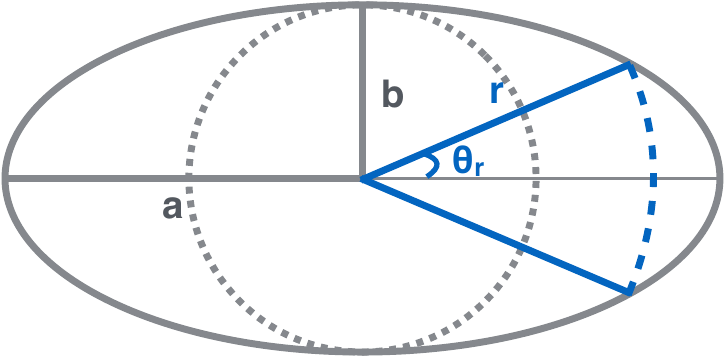}
\caption{Illustration of how cluster ellipticity can lead to a false detection of dependence of
  $\theta_{\rm cen}$ on log($r$/R$_{\rm 200m}$).}
\label{fig:ellipse}
\end{center}
\end{figure}

The result of this process was that satellite-central distance was selected due to the presence of
cluster ellipticity in the model. The reason why adding cluster ellipticity results in the selection
of log($r$/R$_{\rm 200m}$) is illustrated in Fig.~\ref{fig:ellipse}. For satellites with
projected distances $r < b$ (the semi-minor axis of the cluster), the possible values of
$\theta_{\rm cen}$ can vary between 0$^{\circ}$ and 90$^{\circ}$, while for those with $r > b$,
their $\theta_{\rm cen}$ values are confined within $0^{\circ}$ and $\theta{\rm_r}^{\circ} <
90^{\circ}$ due to the boundary of the region contained by the circularized halo radius. Thus, if
satellites were randomly distributed within the elliptical footprint of a cluster, we would expect
that more elliptical clusters exhibit a stronger anti-correlation between log($r$/R$_{\rm 200m}$)
and $\theta_{\rm cen}$, with larger log($r$/R$_{\rm 200m}$) showing smaller $\theta_{\rm cen}$. 
The fact that that anti-correlation is 
not observed suggests that galaxies are not randomly distributed within the elliptical footprint 
of a cluster, but rather are preferentially located on the major axis to a degree that is more 
significant at smaller values of log($r$/R$_{\rm 200m}$). Or viewing in the other way, central galaxies tend
to point toward nearby satellites, whose distribution reflects local, smaller-scale tidal field.

Several previous studies have also investigated the dependence of projected distance on $\theta_{\rm cen}$. 
For isolated host scale, \citet{Brainerd05} found that strength of anisotropy increases with
decreasing projected distance, while \citet{Azzaro07} claimed that the degree of anisotropy is
independent of the projected distance and \citet{Agustsson10} also reported no distance dependence
for red host galaxies. For galaxy group scale, both \citet{Yang06} and \citet{Siverd09} detected
stronger central galaxy alignment effects at smaller projected distance.  Also, in Fig.~2 of
\citet{Faltenbacher07}, $\theta_{\rm cen}$ is smaller in the inner part of halos than in the outer
part for red satellites. 

The general physical picture regarding angular segregation of satellites is that it is due to
large-scale tidal fields, which leads to preferential infall of satellites along the connected filaments. 
This picture is also reflected in the identification of cluster ellipticity as an predictor described in Sec.~\ref{sec: cluster ellipticity}.
If a cluster's small-scale tidal field always followed it's large-scale tidal field, then with the presence of predictor cluster ellipticity (reflecting the direction large-scale tidal field), log($r$/R$_{\rm 200m}$) (reflecting small-scale tidal field) would not be selected out, as all of its effect would be absorbed in cluster ellipticity.
During the chaotic assembly process, a cluster's inner tidal field may differ from it's large-scale tidal field.
The positive correlation between $\theta_{\rm cen}$ and log($r$/R$_{\rm 200m}$) found here implies that 
smaller-scale local tidal field, either newly established or following along the large-scale tidal field, 
does play some role in torquing the central galaxies to align with satellites located relatively nearby as well.

\subsection{Dependence on satellite ellipticity}

We observed a (marginally) statistically significant dependence of the angular segregation of
satellites along the central galaxy major axis on satellite ellipticity, with rounder satellites exhibiting a
stronger tendency to lie along the central galaxy major axis direction.

An intuitive way of interpreting the effect of satellite ellipticity is to link it with related
galaxy properties. Rounder galaxies have less disk component and an older stellar population, and
thus look redder in color. Also, luminous galaxies tend to be rounder in morphology. Given the
relation between satellite ellipticity, color, and luminosity, they may share similar origins, as we
have discussed in Secs.~\ref{subsection: sat color} and~\ref{subsection: sat lum}. However,
notice in Fig.~\ref{fig:dr8_satellite_73146} that the correlation coefficients for satellite
ellipticity with $^{0.1}M_r$ and color are $\sim 0.2$ and $-0.1$ respectively,  meaning that there
exists other physical origins different from the effects of $^{0.1}M_r$ or color. We must seek other physical mechanisms that are more tightly linked to the satellite ellipticity itself.

One possible mechanism for the preference of rounder satellites to lie along the major axis directions of centrals may be their frequent interaction with nearby galaxies.
According to \citet{Kuehn05}, harassment processes due to close encounters with neighboring galaxies
make galaxies rounder. Also, \citet{Rodriguez16} found that elliptical galaxies in groups, where
more disturbing events are likely to happen, are more spherical than field elliptical galaxies with
similar intrinsic properties. Therefore, satellites residing near the major axis directions of centrals
are more likely to be harassed due to the higher number density there. Besides the effect of shaping galaxies, higher frequency interactions with other members let satellites experience through more phase mixing and relation processes, thus speeding up their sinking onto the plane of central galaxy, as the gravitational potential is deeper there.


\section{Summary and conclusion}
\label{sec:summary}

In this work, we investigate the central galaxy alignment effect using the redMaPPer cluster catalog. 
We use three kinds of measurements of the central galaxy position angle from the SDSS derived from previous
work: re-Gaussianization, de Vaucouleurs, and isophotal shapes, compare the derived central galaxy alignment
strength among them, and discuss possible systematic effects. To identify the dominant predictors of
the central galaxy alignment signal, we include as many potential physical parameters as possible, and apply
forward-stepwise linear regression to quantify the statistical significance of these parameters as
predictors, as well as to properly account for correlations between them. 

Our analysis has two steps. In step one, we regress the position angle difference between the central galaxy
and cluster shape (as traced by the member galaxy distribution, a proxy for the dark matter halo
shape), $\Delta \eta$, against central galaxy and cluster related quantities.  The goal of this step is to
identify the central galaxy and cluster properties that most significantly affect their alignment. In step two,
we regress the angular location $\theta_{\rm cen}$ of each member galaxy with reference to its central galaxy
major axis direction against several satellite-related quantities, in order to identify important
predictors for the angular location of the satellite with respect to the central galaxy major axis. Our key
results are as follows.

\begin{enumerate}
\item The detected central galaxy alignment signal is strongest based on isophotal shape, followed by de
  Vaucouleurs and re-Gaussianization shape (see the right panels in Figs.~\ref{fig:Delta_eta_dist}
  and~\ref{fig:BCG_align_ang_dist}). This may be caused by the fact that the isophotal shape traces a galaxy's outermost regions, which are more susceptible to the external tidal fields. 
\item The central galaxy-cluster alignment is strongest for clusters that are more elongated and higher richness, or that have centrals
  with larger physical size, higher luminosity, redder color, and higher centering
  probability\footnote{Although central galaxy ellipticity is found to be a significant predictor as listed in
    Table~\ref{tb:predictor_bcg}, we discussed in Sec.~\ref{sec: BCG e}  that the correlation
    between central galaxy ellipticity and $\Delta \eta$ is more complicated than a simple linear relation, which requires further investigation in future work.}.
\item The tendency of satellites to reside along the central galaxy major axis direction is strongest for
  satellites with redder color, higher luminosity, located closer to its central galaxy 
    and with smaller ellipticity.
\end{enumerate}

As shown, we have selected many predictors that have a statistically significant influence on the
central galaxy alignment effect. This implies that central galaxy alignment is a complicated phenomenon potentially
involved multiple relevant physical processes during galaxy and cluster formation and evolution,
such that it cannot be straightforwardly explained by just few dominant factors. We have discussed
in great detail the potential physical origins of these selected predictors in Secs.~\ref{sec:
  BCG origin} and~\ref{sec: satellite origin}.  The most relevant factors seem to be that central galaxy
alignment may originate from the filamentary accretion processes, but also possibly affected by the
tidal field (either the large-scale primordial tidal field, or  the newly-established small-scale
tidal field after the redistribution of satellites). Also, merger events tend to destroy alignment.
From this work, we cannot fully disentangle the relative contributions from the above three effects,
or rule out contributions from other possible mechanisms that can increase or reduce central galaxy alignment. 
We expect future investigations either based on observations or simulations to put tighter
constraints on possible central galaxy alignment scenarios. 


\section*{Acknowledgements}

We thank Shadab Alam, Tereasa Brainerd, Yun-Hsin Huang, Wentao Luo, Melanie Simet, Sukhdeep Singh, and Ying Zu for useful comments and discussions. 
This work was supported by the National Science Foundation under
Grant No.\ AST-1313169. YC is supported by William S. Dietrich II Presidential Ph.D. Fellowship Award.
EB is partially supported by the US Department of Energy Grant No.\ DE-SC0007901.



\bibliographystyle{mnras}
\bibliography{GAI_reference}  


\bsp	
\label{lastpage}
\end{document}